# Physics-Derived Natural Coordinates for Fast Uncertainty Propagation in Multilayer Thermoreflectance

Tao Chen[1], Dihui Wang[2], Puqing Jiang[1,*]

[1]*School of Power and Energy Engineering, Huazhong University of Science and Technology, Wuhan, Hubei 430074, China*

[2]*Swanson School of Engineering, University of Pittsburgh, Pittsburgh, Pennsylvania, USA*

**ABSTRACT:** Uncertainty quantification in thermoreflectance measurements of multilayer structures is challenging because thermophysical-property extraction involves nonlinear inverse problems with strongly coupled parameters. Covariance-based analytical propagation is efficient but assumes local linearity and approximately Gaussian fitted variables, whereas Monte Carlo (MC) propagation captures non-Gaussian behavior but is computationally expensive and can be unreliable when simultaneously fitting parameters spanning very different orders of magnitude. Here, we address these issues by combining normalized MC sampling, used to obtain reliable benchmark distributions, with physics-based variable transformations for analytical uncertainty propagation. From the multilayer heat-diffusion equation, we identify natural coordinates: the interfacial thermal resistance $1/G$, the substrate combinations $\sqrt{k_z k_r}$ and $\sqrt{k_z C}$, and the thin-film combinations $k_r/C$, $hC$, and $Ch^2/k_z$. Across several representative multilayer systems and thermoreflectance modalities, these coordinates are approximately Gaussian even when the original physical parameters are skewed. Performing covariance-based propagation in the transformed space and mapping the results back to physical parameters yields asymmetric confidence intervals that closely match normalized MC benchmarks. The approach preserves the non-Gaussian uncertainty structure of the recovered parameters while avoiding repeated nonlinear refitting, providing a fast and physically interpretable methodology for thermoreflectance uncertainty quantification in multilayer systems governed by diffusive heat conduction.

**Keywords:** Thermoreflectance; Uncertainty quantification; Monte Carlo analysis; Interfacial thermal conductance; Multilayer structures

*Corresponding Author: jpq2021@hust.edu.cn (P.J.)

## 1. INTRODUCTION

Thermoreflectance techniques have become indispensable non-contact tools in modern thermal metrology for measuring thermal conductivity, volumetric heat capacity, and interfacial thermal conductance in bulk materials, thin films, and multilayer structures [1-4]. Representative methods include time-domain thermoreflectance (TDTR) [5-9], frequency-domain thermoreflectance (FDTR) [10-12], steady-state thermoreflectance (SSTR) [13], spatial-domain thermoreflectance (SDTR) [14], and the recently developed square-pulsed source thermoreflectance (SPS) [15-18]. These methods offer micron-scale spatial resolution, require minimal sample material, and apply to a broad range of micro- and nanoscale heat-transport problems.

At the same time, thermoreflectance measurements are inherently inverse problems: the measured signal must be interpreted through a heat-transfer model to infer the underlying thermophysical parameters. In practice, the sensitivities of the measured signal to different unknown parameters are often strongly coupled, and some parameters may exhibit intrinsically weak sensitivity. Consequently, parameter extraction can become ill-conditioned, leading to substantial uncertainties in the fitted thermophysical properties. Rigorous uncertainty quantification is therefore essential for converting a fitted value into a reliable physical result.

Uncertainty evaluation in thermoreflectance experiments is commonly performed using either covariance-based analytical propagation via local linearization or Monte Carlo (MC) analysis through repeated sampling and nonlinear refitting. In the analytical route, early thermoreflectance studies [13, 19, 20] typically propagated the uncertainties of individual input parameters through first-order sensitivity coefficients, treating the parameters as statistically independent and attributing total uncertainty solely to known inputs, as outlined in the Guide to the Expression of Uncertainty in Measurement (GUM) [21]. This practice effectively neglected two important sources of error: the correlations among fitted parameters introduced by the coupled inverse problem, and the additional uncertainty arising from the quality of the fit itself (i.e., the residual mismatch between the model and the data). Yang *et al*. [22] later derived a more complete analytical formulation that explicitly incorporated both parameter correlations through the full covariance matrix and the contribution of fitting residuals. While this advanced the state of the art, the analytical route still relies fundamentally on local linearization and on the fitted variables being approximately Gaussian. When parameter sensitivities are weak, strongly coupled, or markedly nonlinear, these assumptions may break down, and the analytical results can become inaccurate.

Monte Carlo analysis is often regarded as a more general alternative because it propagates uncertainty through the full nonlinear fitting procedure and can naturally capture parameter coupling and non-Gaussian output distributions. This generality, however, comes at a much higher computational cost. More importantly, its accuracy still depends on whether the fitting algorithm can consistently converge to the global optimum in every trial: improper scaling, unfavorable initial guesses, and convergence to local minima can significantly distort the inferred parameter distributions and thereby compromise the uncertainty estimate [23-25]. Recent efforts to mitigate these robustness issues include deep-learning-assisted fitting to provide better initial guesses [23] and Bayesian parameter estimation (BPE) to unify fitting and uncertainty quantification in a probabilistic framework [25]. While BPE can explicitly account for multivariable correlations and prior knowledge, it generally requires a much broader exploration of parameter space and remains computationally expensive for multilayer systems with several coupled unknowns. Likewise, machine-learning-based approaches can accelerate inference but typically require large training datasets and may lose accuracy outside the training domain.

Existing methods thus involve clear trade-offs between efficiency, generality, robustness, and interpretability. Analytical methods are fast but rely on strong local assumptions; MC methods are more general but computationally demanding and sensitive to fitting robustness; Bayesian and machine-learning approaches offer rigorous or accelerated alternatives but impose their own computational or data requirements. Against this background, two key methodological questions remain unresolved. First, how can we evaluate uncertainty both efficiently and accurately when fitted parameter distributions deviate substantially from Gaussian behavior? Second, under what conditions are Monte Carlo-inferred parameter distributions truly reliable?

In this work, we address both challenges through complementary advances in Monte Carlo sampling and analytical uncertainty propagation. We first establish that parameter normalization, i.e., fitting dimensionless ratios relative to their optimal estimates, is a practical prerequisite for trustworthy MC uncertainty estimation when fitted parameters span vastly different magnitudes. We further demonstrate that thermoreflectance inverse problems possess an intrinsic statistical geometry rooted in the multilayer heat-diffusion equation. This geometry defines natural fitting coordinates where parameter distributions become approximately Gaussian, allowing analytical uncertainty propagation to remain accurate well beyond the locally linear regime. The resulting framework is benchmarked against normalized MC simulations across representative multilayer systems and further validated using different thermoreflectance modalities. These cases encompass (i) metal-dielectric interfaces with weak interfacial sensitivity, (ii) anisotropic substrates where the in-plane and cross-plane conductivities are strongly coupled, and (iii) multilayer thin films with simultaneous uncertainty in thickness, thermal conductivity, and volumetric heat capacity. Close agreement between the transformed analytical and MC results is obtained in all cases, confirming that the analytical route retains the statistical fidelity of the MC approach at a fraction of the computational cost.

## 2. METHODOLOGIES

### *2.1 Inverse problem formulation*

Thermoreflectance experiments probe the thermal response of multilayer structures under periodic optical heating. As illustrated in Fig. 1(a), the sample is modeled as a multilayer stack consisting of a metal transducer, one or more thin films, and a substrate. Each layer is characterized by its in-plane thermal conductivity $k_r$, cross-plane thermal conductivity $k_z$, volumetric heat capacity $C$, and thickness $h$, while each interface is described by an interfacial thermal conductance $G$. The laser spot size $r_0$ is an additional geometric input. In practice, some quantities are independently measured or otherwise known, whereas the remaining quantities constitute the unknown parameter vector $\mathbf{u}$ to be determined. Here, FDTR is adopted as a representative technique to illustrate the uncertainty-analysis framework. Its applicability to other thermoreflectance modalities is further examined in Section S3 of the Supplementary Information. Figure 1(b) shows experimental FDTR signals measured from an Al/silica sample, together with the corresponding thermal model predictions.

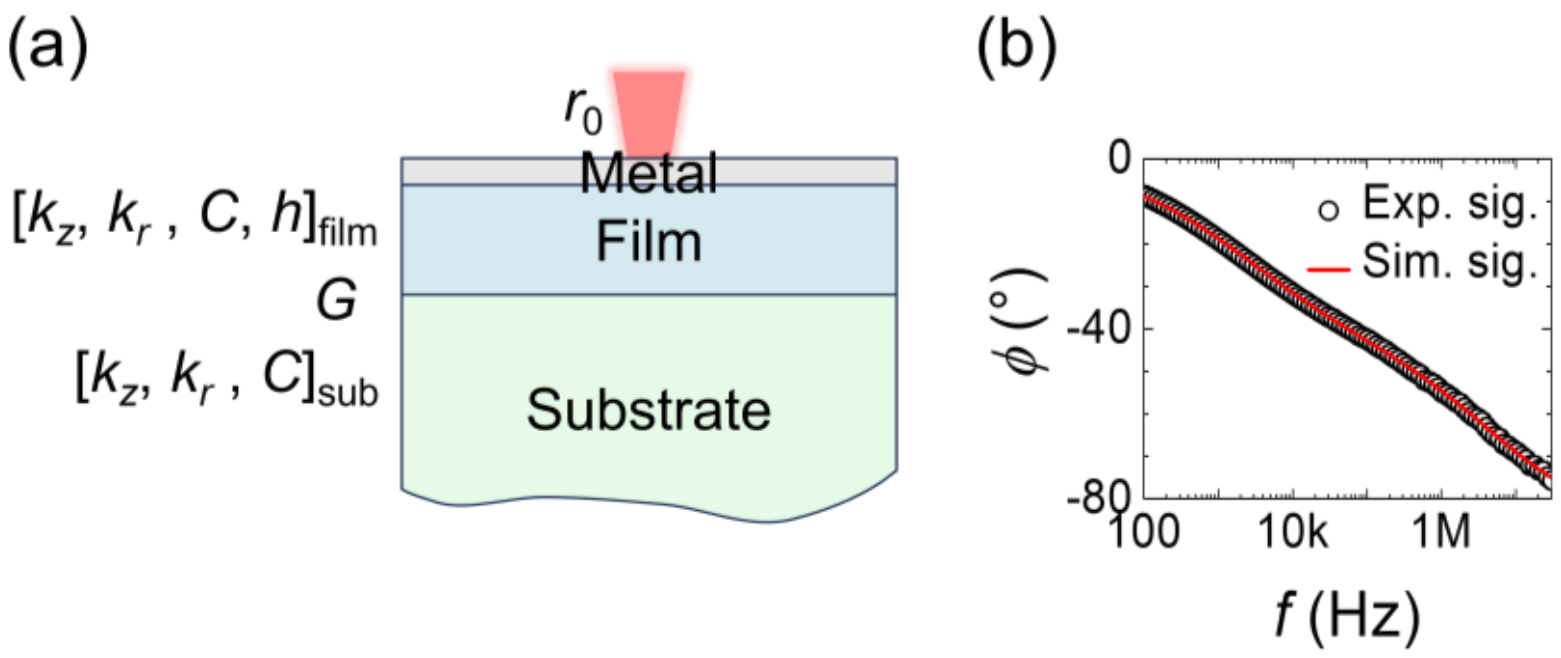

**FIG. 1.** Multilayer sample geometry and inverse-problem schematic: (a) multilayer structure consisting of metal transducer, thin film(s), and substrate; (b) comparison between measured FDTR signals and model predictions for extracting unknown thermophysical parameters.

Mathematically, the measured signal can be expressed as

$$y_{\text{exp}} = y_{\text{sim}}(\mathbf{p}, \mathbf{u}) + \varepsilon \tag{1}$$

where $\mathbf{p}$ denotes the vector of known input parameters, $\mathbf{u}$ the vector of unknown parameters, and $\varepsilon$ the experimental noise. The forward model $y_{\text{sim}}$ is obtained from the heat diffusion equation in cylindrical coordinates for a transversely isotropic medium

$$C\frac{\partial T}{\partial t} = \frac{k_r}{r}\frac{\partial}{\partial r}\left(r\frac{\partial T}{\partial r}\right) + k_z\frac{\partial^2 T}{\partial z^2} \tag{2}$$

where $T$ is the temperature rise. At the interface between two adjacent layers, the temperature discontinuity due to finite interfacial thermal conductance is

$$T_1 - T_2 = q\frac{1}{G} \tag{3}$$

with $q$ the interfacial heat flux.

Applying Hankel and Fourier transforms recasts the multilayer heat-conduction problem in transfer-matrix form. The detector signal is then calculated by incorporating the temporal profile of the heat source and the Gaussian spatial weighting of the pump and probe beams. Although specific thermoreflectance techniques differ in heating waveforms and acquisition schemes, their inverse problems share the same essential structure: the measured thermal response is compared with a forward model to extract $\mathbf{u}$. This process is illustrated schematically in Fig. 1(b), where the mismatch between measured and simulated signals is minimized to obtain the best-fit parameters.

For $M$ simultaneously fitted datasets, the objective function is

$$J(\mathbf{u};\mathbf{p}) = \prod_{j=1}^{M} \mathrm{MSD}_j \tag{4}$$

where $\mathrm{MSD}_j$ is the mean squared deviation between the $j$-th measured signal and its simulated counterpart. The optimal parameter set $\hat{\mathbf{u}}$ is therefore defined by

$$\hat{\mathbf{u}} = \arg\min_{\mathbf{u}} J(\mathbf{u};\mathbf{p}) \tag{5}$$

This inverse-fitting step is the origin of uncertainty propagation, because the fitted results depend on both measurement noise and uncertainties in the known input $\mathbf{p}$.

### *2.2 Monte Carlo uncertainty propagation*

A direct way to evaluate uncertainty is the Monte Carlo approach, in which the known input parameters $\mathbf{p}$ are treated as random variables rather than exact constants. Each $p_i$ is assigned a probability distribution—typically Gaussian, $p_i \sim \mathcal{N}(p_i^0, \sigma_i^2)$, where $p_i^0$ is the nominal value and $\sigma_i$ is the standard uncertainty. In each of $N_{\mathrm{MC}}$ trials, a realization $\mathbf{p}^{(j)}$ is sampled from the prescribed input distributions.

The experimental signal is also independently perturbed by additive Gaussian noise. When repeated measurements are available, the noise statistics can be determined directly at each data point and used to generate the perturbed signals. In the absence of repeated measurements, the best-fit simulated signal evaluated at the nominal parameter values is used as an approximation to the underlying noise-free signal,

$$\mathbf{y}_{\text{exp}}^{(j)} = \mathbf{y}_{\text{sim}}^{0} + \boldsymbol{\varepsilon}^{(j)}, \boldsymbol{\varepsilon}^{(j)} \sim \mathcal{N}\left(\mathbf{0}, \sigma_{\text{exp}}^{2}\mathbf{I}\right) \tag{6}$$

where $\mathbf{y}_{\text{sim}}^{0} = f(\mathbf{p}^{0}, \mathbf{u}^{0})$. The same noise standard deviation $\sigma_{\text{exp}}$ is assumed for all data points and is estimated from the root-mean-square noise across all data points.

The perturbed signal $\mathbf{y}_{\text{exp}}^{(j)}$ is then refitted using the sampled known parameters, the forward model, and the optimization criterion of Eq. (5) to obtain a new estimate $\hat{\mathbf{u}}^{(j)}$. The ensemble of fitted estimates yields the statistical distribution of the unknown parameters. The workflow is schematically illustrated in Fig. 2.

Monte Carlo analysis naturally captures parameter coupling, nonlinear model response, and non-Gaussian output distributions. However, it is not automatically reliable. When the fitted parameters differ greatly in magnitude, direct fitting in the original dimensional space produces a poorly scaled optimization landscape: convergence tolerances and step-size controls that are appropriate for one parameter become ill-suited for another, increasing the risk of convergence to local minima or premature termination. The resulting inferred distributions are artificially broadened, and the uncertainty is overestimated.

To avoid this problem, we normalize the fitting variables:

$$u_i' = \frac{u_i}{u_i^0} \tag{7}$$

where $u_i^0$ is the optimal reference value obtained from the nominal fit. All normalized variables remain near unity, ensuring a balanced parameter space in which the optimizer operates under uniform relative tolerances. After fitting, the physical distributions are recovered by multiplying the fitted $u_i'$ by $u_i^0$. As demonstrated in Section 3.1, this normalization is a prerequisite for trustworthy MC uncertainty estimates when the fitted parameters span disparate orders of magnitude.

Despite its generality, MC analysis remains computationally expensive because each trial requires a complete nonlinear fit. For multilayer models with several coupled unknowns, wide frequency ranges, or multiple simultaneously fitted datasets, the repeated optimization can become prohibitively time-consuming. This limitation motivates the analytical alternative developed next.

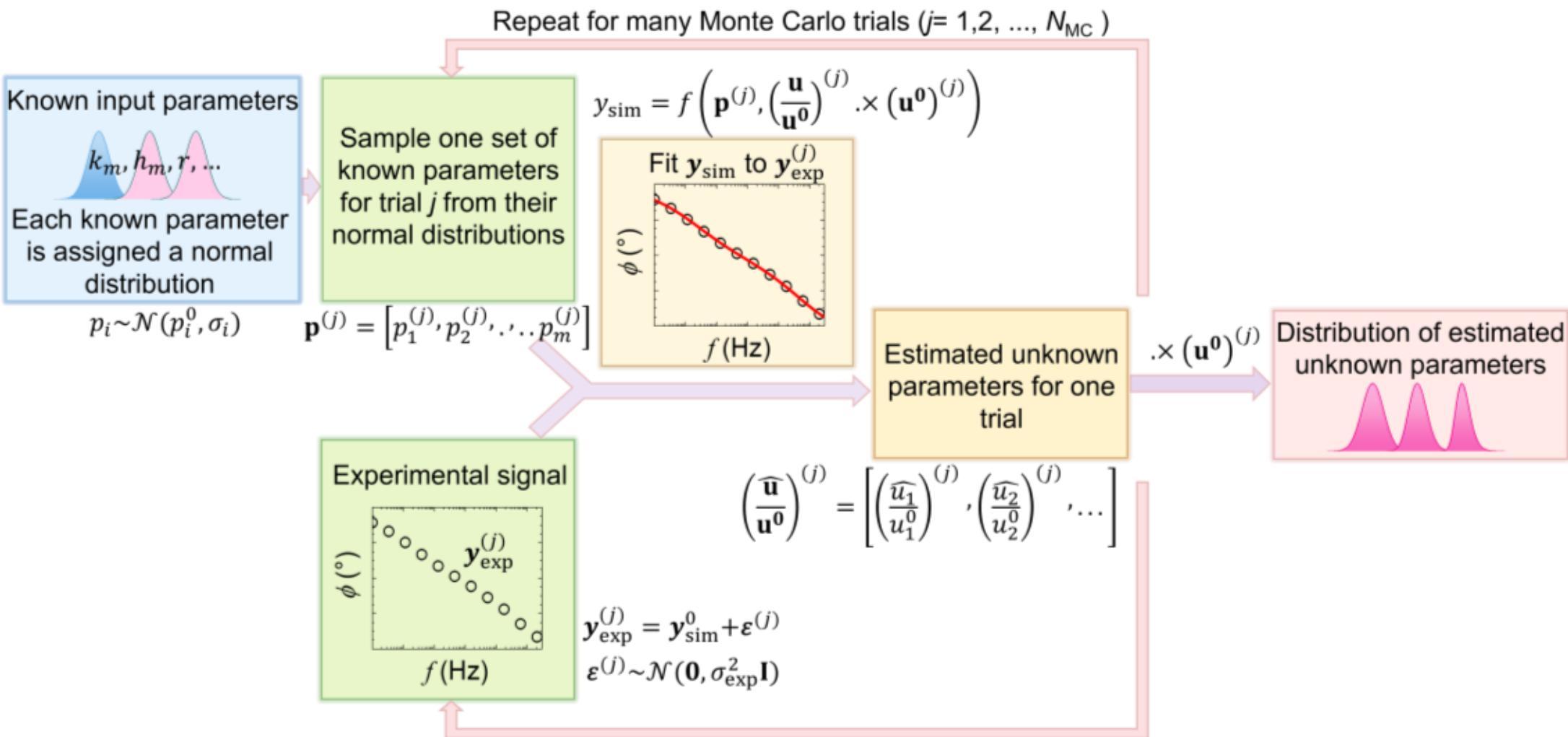

**FIG. 2.** Schematic workflow of Monte Carlo uncertainty quantification with normalized fitting. In each trial, the known input parameters are randomly sampled from their prescribed distributions, while the experimental signal is independently perturbed by additive Gaussian noise with the same standard deviation for all data points. The sampled inputs and perturbed signal are then used for refitting in the normalized parameter space ($u_i' = u_i/u_i^0$), and the fitted result is recovered to physical units. Repeating this procedure yields the statistical distribution of the fitted parameters.

### *2.3 Analytical uncertainty propagation*

The covariance-based formulation below follows Yang et al. [22] and is summarized only briefly to fix notation. The analytical method estimates parameter uncertainties by locally linearizing the inverse problem around the best-fit solution. For the $j$-th dataset, let $y_j(x_i)$ denote the measured signal at point $x_i$ and $g_j(\mathbf{u}, \mathbf{p}, x_i)$ the model prediction. The mean squared deviation is

$$\mathrm{MSD}_j = \frac{1}{N_j} \sum_{i=1}^{N_j} \left[ g_j(\mathbf{u}, \mathbf{p}, x_i) - y_j(x_i) \right]^2 \tag{8}$$

and the global objective function is written as $J(\mathbf{u},\mathbf{p}) = \prod_{j=1}^{M} \mathrm{MSD}_j$, where $N_j$ is the number of data points in the $j$-th dataset. At the optimal fit, the gradient of $J$ with respect to each unknown parameter is zero. Expanding the stationarity condition to the first order about $(\mathbf{u}^0, \mathbf{p}^0)$ gives

$$g_j(\hat{\mathbf{u}}, \tilde{\mathbf{p}}, x_i) \approx g_j(\mathbf{u}^0, \mathbf{p}^0, x_i) + \sum_{l=1}^{n_u} \left.\frac{\partial g_j}{\partial u_l}\right|_{(\mathbf{u}^0,\mathbf{p}^0,x_i)} (\hat{u}_l - u_l^0) + \sum_{k=1}^{n_p} \left.\frac{\partial g_j}{\partial p_k}\right|_{(\mathbf{u}^0,\mathbf{p}^0,x_i)} (\tilde{p}_k - p_k^0) \quad (9)$$

where $\tilde{\mathbf{p}}$ is a perturbed realization of the known inputs and $\hat{\mathbf{u}}$ the corresponding fitted parameters.

In vector form, the measured and modeled signals for the $j$-th dataset are

$$\mathbf{e}_j = \left[y_j(x_1), y_j(x_2), \dots, y_j(x_{N_j})\right]^{\mathrm{T}}$$

$$\mathbf{f}_j = \left[g_j(\mathbf{u}^0, \mathbf{p}^0, x_1), g_j(\mathbf{u}^0, \mathbf{p}^0, x_2), \dots, g_j(\mathbf{u}^0, \mathbf{p}^0, x_{N_j})\right]^{\mathrm{T}} \quad (10)$$

The Jacobian matrices are

$$\mathbf{J}_{u,j} = \left[\frac{\partial \mathbf{f}_j}{\partial u_1}, \frac{\partial \mathbf{f}_j}{\partial u_2}, \dots, \frac{\partial \mathbf{f}_j}{\partial u_{n_u}}\right], \mathbf{J}_{p,j} = \left[\frac{\partial \mathbf{f}_j}{\partial p_1}, \frac{\partial \mathbf{f}_j}{\partial p_2}, \dots, \frac{\partial \mathbf{f}_j}{\partial p_{n_p}}\right] \quad (11)$$

and the diagonal weighting matrix accounting for normalized residuals is

$$\mathbf{G}_j = \mathrm{diag}\left[\frac{1}{y_j^2(x_1)}, \frac{1}{y_j^2(x_2)}, \dots, \frac{1}{y_j^2(x_{N_j})}\right] \quad (12)$$

Substituting the linearized response into the stationarity condition and neglecting higher-order terms yields the parameter perturbation

$$\hat{\mathbf{u}} - \mathbf{u}^0 = \Sigma_{uu}^{-1}\mathbf{b} - \Sigma_{uu}^{-1}\Sigma_{up}(\tilde{\mathbf{p}} - \mathbf{p}^0) \quad (13)$$

provided that $\Sigma_{uu}$ is non-singular, where

$$\Sigma_{uu} = \sum_{j=1}^{M} \alpha_j\, \mathbf{J}_{u,j}^{\mathrm{T}} \mathbf{G}_j \mathbf{J}_{u,j}$$

$$\Sigma_{up} = \sum_{j=1}^{M} \alpha_j\, \mathbf{J}_{u,j}^{\mathrm{T}} \mathbf{G}_j \mathbf{J}_{p,j}$$

$$\mathbf{b} = \sum_{j=1}^{M} \alpha_j\, \mathbf{J}_{u,j}^{\mathrm{T}} \mathbf{G}_j (\mathbf{e}_j - \mathbf{f}_j)$$

and

$$\alpha_j = \frac{1}{N_j} \prod_{k \neq j} \mathrm{MSD}_k \tag{14}$$

Assuming that the signal residuals and the uncertainties in the known parameters are mutually independent, the covariance matrix of the fitted parameters is

$$\mathrm{Var}[\hat{\mathbf{u}}] = \Sigma_{uu}^{-1} \left[ \sum_{j=1}^{M} \alpha_j^2 \, \mathbf{J}_{u,j}^{\mathrm{T}} \mathbf{G}_j \, \mathrm{Var}[\mathbf{e}_j - \mathbf{f}_j] \, \mathbf{G}_j^{\mathrm{T}} \mathbf{J}_{u,j} \right] \Sigma_{uu}^{-1} + \Sigma_{uu}^{-1} \Sigma_{up} \, \mathrm{Var}[\tilde{\mathbf{p}}] \, \Sigma_{up}^{\mathrm{T}} \Sigma_{uu}^{-1} \tag{15}$$

The first term represents uncertainty from experimental noise and fitting residuals; the second term represents the uncertainty propagated from the known inputs. In practice, $\mathrm{Var}[\mathbf{e}_j - \mathbf{f}_j]$ is estimated from the fitting residuals of the $j$-th dataset, whereas $\mathrm{Var}\,[\tilde{\mathbf{p}}]$ is constructed from the variances and, when applicable, covariances of the known parameters. For a non-diagonal input covariance matrix, we write $\mathrm{Var}\,[\tilde{\mathbf{p}}] = \mathbf{M} + \mathbf{N}$, where $\mathbf{M}$ contains the individual input variances and $\mathbf{N}$ the off-diagonal covariances. The second term in Eq. (15) then contains an additional contribution, $\Sigma_{uu}^{-1}\Sigma_{up}\mathbf{N}\Sigma_{up}^{\mathrm{T}}\Sigma_{uu}^{-1}$, which may increase or decrease the variance of an individual fitted parameter and may also modify the correlations among fitted parameters.

The covariance matrix

$$\mathrm{Var}[\hat{\mathbf{u}}] = \begin{bmatrix} \sigma_{u_1}^2 & \mathrm{cov}(u_1, u_2) & \cdots \\ \mathrm{cov}(u_2, u_1) & \sigma_{u_2}^2 & \cdots \\ \vdots & \vdots & \ddots \end{bmatrix} \tag{16}$$

provides the variances (diagonal) and covariances (off-diagonal). The standard uncertainty of each fitted parameter is obtained as the square root of the corresponding diagonal element of $\mathrm{Var}[\mathbf{u}]$. All uncertainties in this work are reported as expanded uncertainties at coverage factor $k = 2$, corresponding to an approximate 95.45% coverage interval under the assumption of normality.

The above formulation is computationally efficient but relies on the fitted parameters being approximately Gaussian in the original space—a condition frequently violated in thermoreflectance. We therefore introduce a transformed variable space

$$\mathbf{z} = \boldsymbol{\Phi}(\mathbf{u}) \tag{17}$$

where $\boldsymbol{\Phi}(\cdot)$ denotes physically motivated transformations. Analytical propagation is performed with respect to $\mathbf{z}$, yielding $\mathrm{Var}[\hat{\mathbf{z}}]$, and the uncertainties of the original physical parameters are recovered through the inverse mapping $\mathbf{u} = \Phi^{-1}(\mathbf{z})$. Throughout this work, we use the term "natural coordinates" to denote the parameter coordinates in which the fitted distributions are approximately Gaussian. As shown in the next section, these coordinates are not the original physical parameters themselves but specific reciprocal, square-root, or composite combinations of them, emerging from the structural form of the multilayer heat-diffusion equation.

### *2.4 Physics-motivated variable transformation*

The transformation anticipated in Section 2.3 follows directly from the mathematical structure of the multilayer heat-diffusion problem. Rather than postulating these coordinates empirically, we derive them from the governing equation to establish an a priori expectation for the natural statistical coordinates of the inverse problem.

Consider first the interfacial thermal conductance. Equation (3) shows that, for a given interfacial heat flux $q$, the temperature discontinuity across the interface is strictly linear in the interfacial thermal resistance $1/G$. Because thermoreflectance is fundamentally a thermometry technique that measures temperature response rather than heat flux directly, $1/G$ is the quantity that enters the boundary condition linearly. Consequently, from the perspective of the temperature field, $1/G$ is the more natural variable than $G$ itself.

For the substrate, we apply the Fourier transform to the time variable $t$ and the Hankel transform to the radial coordinate $r$ in Eq. (2). The governing equation then reduces to an ordinary differential equation in the depth coordinate,

$$\frac{\partial^2 \Theta}{\partial z^2} = \lambda^2 \Theta \tag{18}$$

where $\Theta$ is the transformed temperature, $\lambda = \sqrt{4\pi^2\rho^2 k_r/k_z + i\omega C/k_z}$, $\rho$ is the Hankel transform variable, and $\omega$ is the angular frequency. The general solution of Eq. (18) is $\Theta = e^{\lambda z}B^+ + e^{-\lambda z}B^-$, and the heat flux $Q = -k_z(\mathrm{d}\Theta/\mathrm{d}z)$ becomes

$$Q = \gamma\left(-e^{\lambda z}B^+ + e^{-\lambda z}B^-\right) \tag{19}$$

where $B^+$ and $B^-$ are complex constants determined by the boundary conditions, and

$$\gamma = k_z\lambda = \sqrt{4\pi^2\rho^2(k_r k_z) + i\omega k_z C} \tag{20}$$

For a semi-infinite substrate, finiteness as $z \to \infty$ requires $B^+_{\mathrm{sub}} = 0$, yielding at the substrate surface ($z = 0$)

$$\Theta_{\mathrm{sub},z=0} = B^-_{\mathrm{sub}}, \qquad Q_{\mathrm{sub},z=0} = \gamma B^-_{\mathrm{sub}} = \gamma\Theta_{\mathrm{sub},z=0} \tag{21}$$

Thus, the substrate enters the boundary-value problem exclusively through the single composite parameter $\gamma$. Inspection of Eq. (20) reveals that $\gamma$ is governed by two characteristic combinations: in the low-frequency limit ($\omega \to 0$), the real part $4\pi^2\rho^2(k_r k_z)$ dominates, so the relevant combination is the geometric-mean thermal conductivity $\sqrt{k_r k_z}$; in the high-frequency limit, the imaginary part $\omega k_z C$ dominates, and the relevant combination is the thermal effusivity $\sqrt{k_z C}$. These combinations therefore constitute the natural coordinates for substrate-related properties.

The situation differs for a finite film of thickness $h$. In this case, the temperature response is governed not only by $\gamma$ but also by the dimensionless product $\lambda h$. From the definition of $\lambda$, there is

$$\lambda h = \sqrt{4\pi^2\rho^2\frac{k_r}{C}\frac{Ch^2}{k_z} + i\omega\frac{Ch^2}{k_z}} \tag{22}$$

This expression shows that the spatial evolution of temperature within the film is controlled by the in-plane thermal diffusivity $k_r/C$ and the cross-plane diffusion time $Ch^2/k_z$. In addition, the amplitude of the film temperature rise is governed by the areal heat capacity $hC$. Therefore, the natural thin-film variables are the thermal diffusivity $k_r/C$, the areal heat capacity $hC$, and the cross-plane diffusion time $Ch^2/k_z$.

These physical considerations suggest that the approximately Gaussian variables in thermoreflectance inverse problems are:

- **interfacial transport:** the thermal resistance $1/G$;
- **substrate properties:** the geometric-mean conductivity $\sqrt{k_r k_z}$ and the thermal effusivity $\sqrt{k_z C}$;
- **thin-film properties:** the in-plane thermal diffusivity $k_r/C$, the areal heat capacity $hC$, and the cross-plane diffusion time $Ch^2/k_z$.

Whether these coordinates indeed yield approximately normal distributions under realistic measurement conditions is not guaranteed by the linear structure of the governing equation alone, because the inverse problem remains nonlinear and the sensitivity landscape depends on the specific experimental configuration. Their validity is therefore examined numerically in Section 3.

## 3. RESULTS AND DISCUSSION

To systematically examine the statistical structure of thermoreflectance inverse problems, we investigate six representative multilayer systems spanning interfacial, substrate, and thin-film parameter categories. Figure 3 illustrates the layer stacks: (a) Al–silica, (b) Al–500 nm $SiO_2$–Si, (c) Al–320 nm GaO–SiC, (d) Al–2 μm $SiO_2$–Si, (e) Al–1.08 μm GaN–0.46 μm AlGaN–Si, and (f) Al–1.59 μm Si–1.03 μm $SiO_2$–Si. The frame color of each panel indicates the parameter category under investigation: red for interfacial transport, blue for substrate properties, and green for thin-film properties. Within each schematic, unknown fitting parameters are highlighted in bold red font, whereas known parameters are shown in black.

Before examining the distribution types of fitted parameters, we first establish a numerical prerequisite for trustworthy Monte Carlo estimation. Section 3.1 uses the Al–GaN–AlGaN–Si system [Fig. 3(e)] to demonstrate that parameter normalization is essential when fitted parameters differ greatly in magnitude.

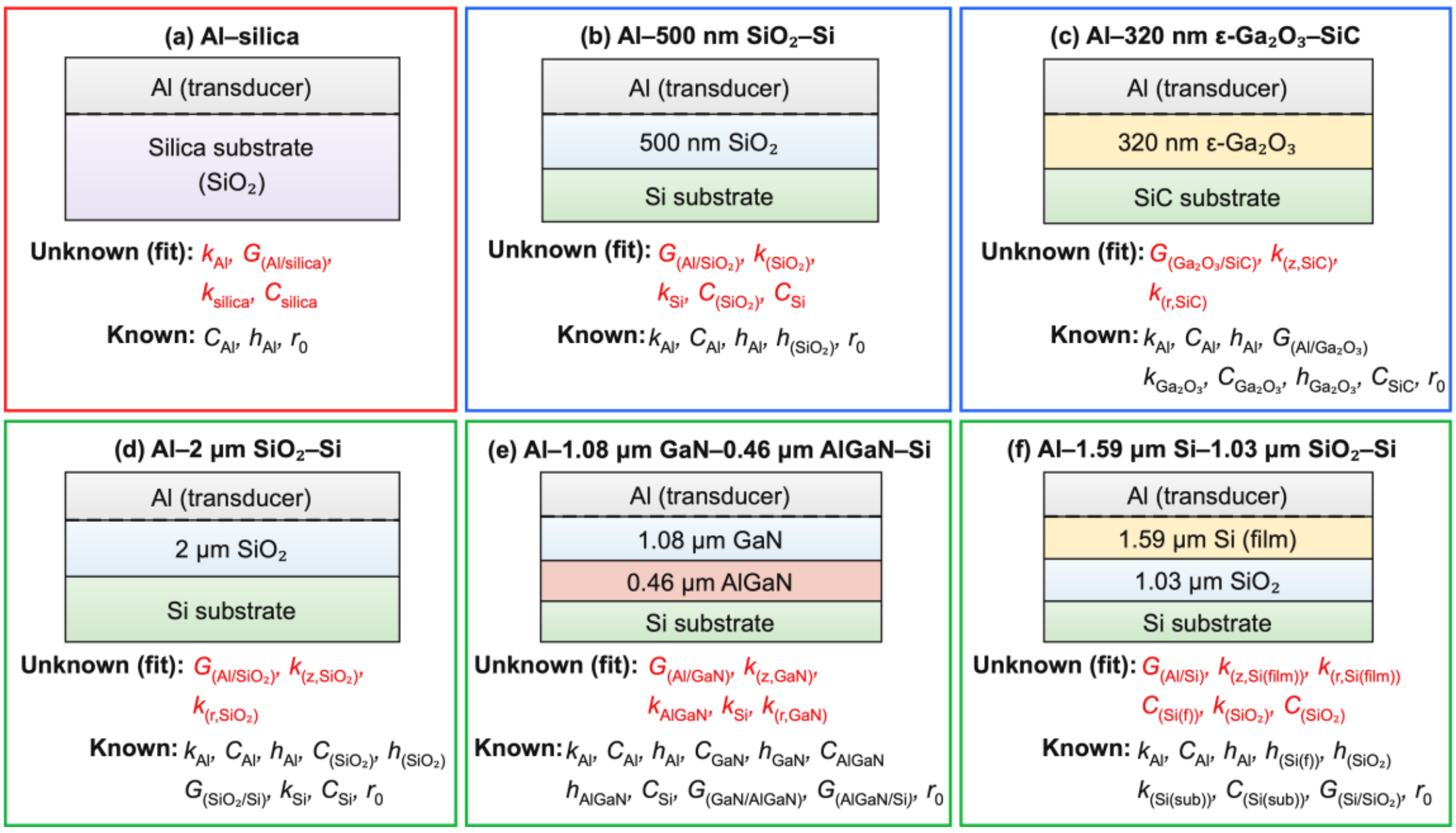

**FIG. 3.** Schematic of the six multilayer systems investigated in this work: (a) Al–silica, (b) Al–500 nm $SiO_2$–Si, (c) Al–320 nm GaO–SiC, (d) Al–2 μm $SiO_2$–Si, (e) Al–1.08 μm GaN–0.46 μm AlGaN–Si, and (f) Al–1.59 μm Si–1.03 μm $SiO_2$–Si. The frame color of each panel indicates the parameter category under investigation: red for interfacial thermal conductance, blue for substrate properties, and green for thin-film properties. Unknown fitting parameters are highlighted in bold red font; known parameters are shown in black.

### *3.1 Normalized Monte Carlo as a numerical prerequisite*

The Al–1.08 μm GaN–0.46 μm AlGaN–Si system [Fig. 3(e)] is used to establish this numerical prerequisite. The measured FDTR signals are taken from Ref. [26]. Five parameters are treated as unknown: the interfacial thermal conductance $G_{\mathrm{Al/GaN}}$, the cross-plane thermal conductivity of GaN film $k_{z,\mathrm{GaN}}$, the thermal conductivity of AlGaN film $k_{\mathrm{AlGaN}}$, the thermal conductivity of Si substrate $k_{\mathrm{Si}}$, and the in-plane thermal conductivity of GaN film $k_{r,\mathrm{GaN}}$. The known parameters include the Al transducer properties ($k_{\mathrm{Al}}$, $C_{\mathrm{Al}}$, $h_{\mathrm{Al}}$), the GaN and AlGaN thicknesses and heat capacities ($C_{\mathrm{GaN}}$, $h_{\mathrm{GaN}}$, $C_{\mathrm{AlGaN}}$, $h_{\mathrm{AlGaN}}$), the Si substrate heat capacity ($C_{\mathrm{Si}}$), the deeper interfacial conductances ($G_{\mathrm{GaN/AlGaN}}$, $G_{\mathrm{AlGaN/Si}}$), and the laser spot size ($r_0$). Their uncertainties are set to 5% for thermal conductivities, 3% for volumetric heat capacities, 2% for thicknesses and $r_0$, 50% for $G_{\mathrm{GaN/AlGaN}}$, and 25% for $G_{\mathrm{AlGaN/Si}}$.

To reduce the risk of convergence to local optima across thousands of MC trials, a hybrid optimization strategy [27] is adopted. A particle swarm optimization (PSO) algorithm first searches the parameter space globally with a swarm size of 200 and 20 generations. Once the global optimum region is identified, a quasi-Newton local optimizer refines the solution. This strategy is far less sensitive to the initial guess than conventional local optimization alone, because the first PSO generation already provides broad coverage of the admissible parameter space. Broad fitting bounds were used for all unknown parameters to minimize the influence of parameter constraints on the resulting MC distributions. Depending on parameter sensitivity, the bounds were set to 0.1–10, 0.2–5, or 0.5–2 times the corresponding nominal values, all of which extend well beyond the ranges sampled by the resulting MC distributions. Under the computational settings used here, one MC trial required approximately 3 s on average.

Figure 4 compares the MC distributions obtained with and without parameter normalization. Figures 4(a–e) show the results obtained by directly fitting the original dimensional parameters. Only $G_{\mathrm{Al/GaN}}$ exhibits an approximately normal distribution, whereas the other four parameters show pronounced skewness and substantially broader distributions. By contrast, Figs. 4(f–j) show the recovered distributions obtained from normalized fitting, as defined in Section 2.2. In principle, because normalization is a one-to-one reparameterization, the two approaches should yield identical results in the physical parameter space if every fit converges accurately to the same global optimum. A trial-by-trial comparison of the final objective-function values further shows that normalized fitting yields lower values in all MC trials, while the corresponding root-mean-square fitting residuals are comparable to the experimental noise level (Section S1, Supplementary Information). The observed discrepancy therefore reflects numerical optimization effects.

Because the fitted parameters span several orders of magnitude, direct fitting in the original dimensional parameter space produces a poorly scaled optimization landscape, for which convergence tolerances appropriate for one parameter may be unsuitable for another, leading to inconsistent convergence and broader MC distributions. Parameter normalization improves the scaling of the optimization problem and promotes more consistent convergence. For the present problem, normalization is therefore not merely a numerical convenience but a practical prerequisite for trustworthy MC uncertainty estimation when the fitted parameters span several orders of magnitude.

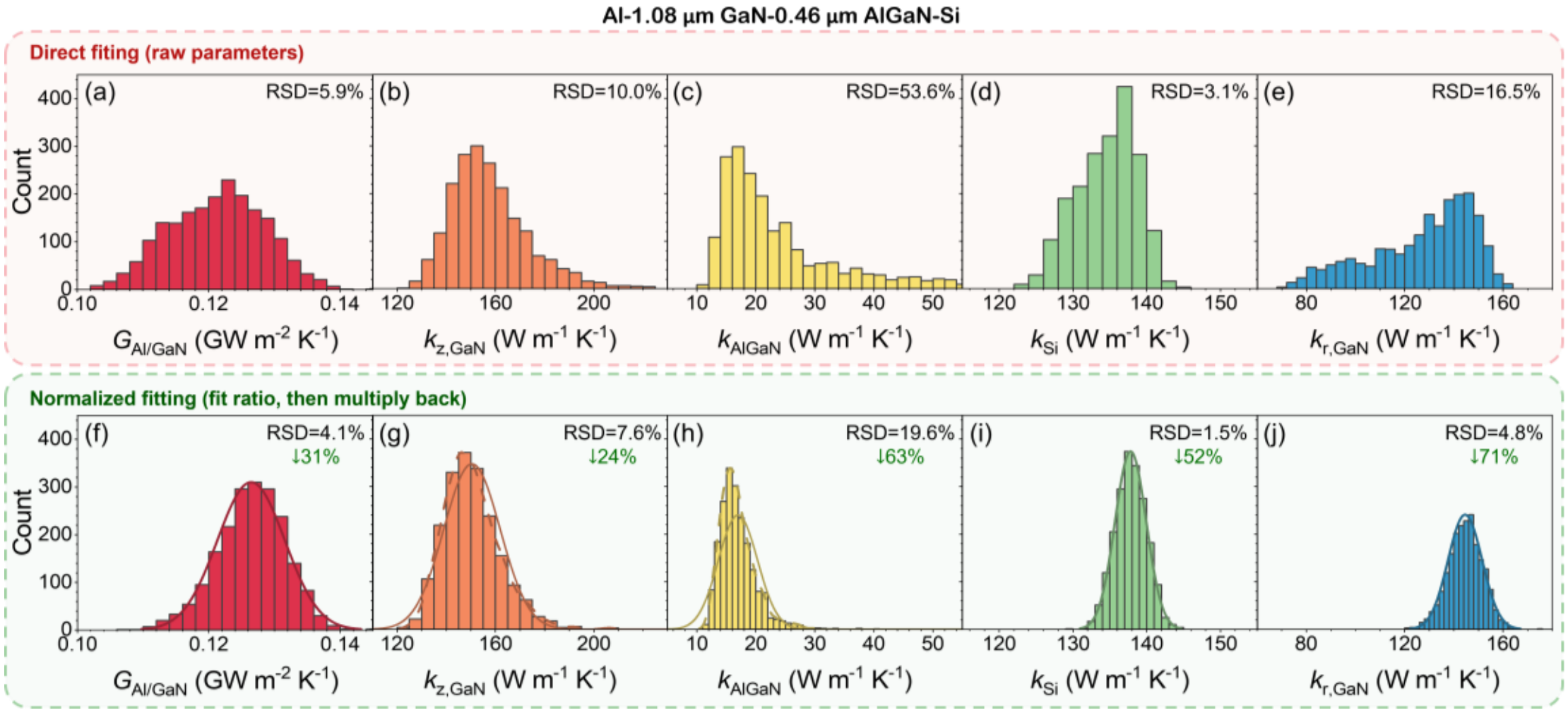

**FIG. 4.** Monte Carlo distributions of the five fitted parameters for the Al–GaN–AlGaN–Si case. (a–e) Distributions obtained by fitting the original dimensional parameters directly. (f–j) Distributions of the recovered physical parameters obtained from normalized fitting. The relative standard deviation (RSD) annotated in each panel quantifies the distribution width; the percentage reduction (green) indicates the uncertainty contraction achieved by normalization. The solid lines represent Gaussian fitting curves, and the dashed lines represent kernel-smoothed curves.

All Monte Carlo analyses presented in the following sections are therefore performed in the normalized parameter space. With this numerical foundation established, we now turn to the central question: in what natural coordinates are the fitted parameters of thermoreflectance inverse problems approximately Gaussian?

*3.2 Reciprocal-normal variable for interfacial transport*

We first test the theoretical expectation that interfacial thermal transport is naturally described by the interfacial thermal resistance $1/G$ rather than the conductance $G$. The Al–silica system [Fig. 3(a)] is used for this purpose. The known parameters are the Al transducer heat capacity $C_{\text{Al}}$, thickness $h_{\text{Al}}$, and laser spot size $r_0$, with relative uncertainties of 3%, 2%, and 2%, respectively. The unknown fitting parameters are the Al thermal conductivity $k_{\text{Al}}$, the interfacial thermal conductance $G_{\text{Al/silica}}$, the silica thermal conductivity $k_{\text{silica}}$, and the silica heat capacity $C_{\text{silica}}$.

Figure 5 shows the Monte Carlo distribution obtained by fitting these original variables. Among the four parameters, $G_{\text{Al/silica}}$ exhibits a clearly right-skewed distribution [Fig. 5(b)], whereas $k_{\text{Al}}$, $k_{\text{silica}}$, and $C_{\text{silica}}$ remain comparatively symmetric [Fig. 5(a), 5(c), and 5(d)]. Under such conditions, conventional analytical uncertainty propagation, which produces a symmetric interval around the fitted value, becomes inadequate, because the Monte Carlo result indicates a distinctly asymmetric distribution with a larger positive deviation and a smaller negative deviation.

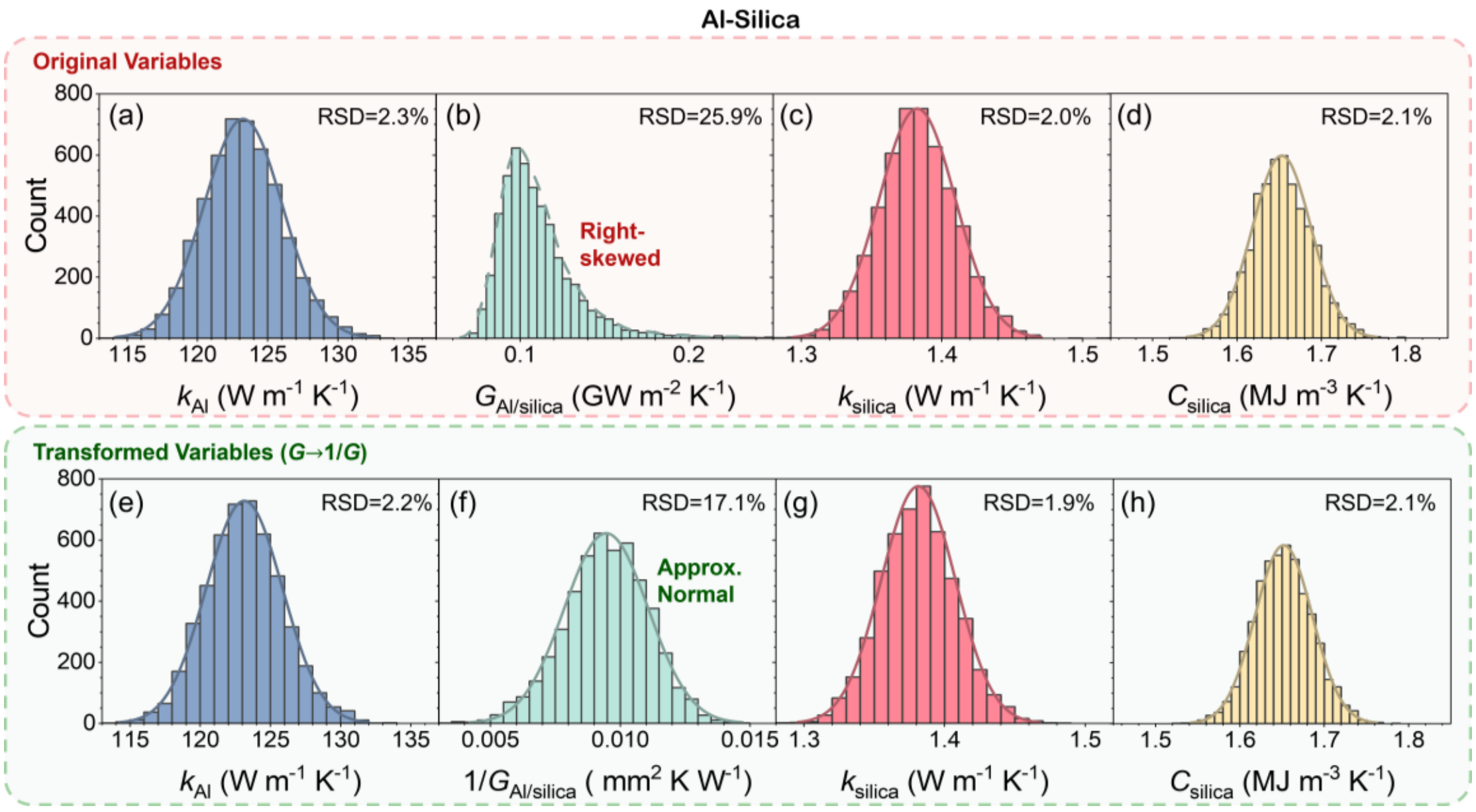

**FIG. 5.** Monte Carlo distributions for the Al–silica case. (a–d) Distributions obtained by fitting the original variables $k_{\text{Al}}$, $G_{\text{Al/silica}}$, $k_{\text{silica}}$, and $C_{\text{silica}}$. (e–h) Distributions obtained after replacing $G_{\text{Al/silica}}$ with its reciprocal $1/G_{\text{Al/silica}}$. The right-skewed distribution of $G_{\text{Al/silica}}$ in (b) becomes approximately normal in (f), while the other parameters remain statistically unchanged, confirming that $1/G$ is the natural coordinate for interfacial thermal transport.

To examine the origin of this asymmetry, the fitting variables were redefined by replacing $G_{\text{Al/silica}}$ with its reciprocal $1/G_{\text{Al/silica}}$, while keeping $k_{\text{Al}}$, $k_{\text{silica}}$, and $C_{\text{silica}}$ unchanged. The corresponding MC distributions are shown in Fig. 5(e–h). A direct comparison reveals that the distributions of $k_{\text{Al}}$, $k_{\text{silica}}$, and $C_{\text{silica}}$ remain almost unchanged. This confirms that transforming one fitting parameter does not materially alter the statistical behavior of the other uncoupled parameters. More importantly, the transformed variable $1/G_{\mathrm{Al/silica}}$ now exhibits an approximately normal distribution [Fig. 5(f)], validating the theoretical prediction of Section 2.4.

This result provides a natural way to apply the covariance-based analytical uncertainty method. Since $1/G_{\mathrm{Al/silica}}$ is approximately Gaussian, the analytical confidence interval can be evaluated in the reciprocal coordinate and then mapped back to the conductance coordinate. Because $G = 1/(1/G)$, the lower bound of $1/G$ corresponds to the upper bound of $G$, while the upper bound of $1/G$ corresponds to the lower bound of $G$. Let $\bar{x}$ and $2\sigma_x$ denote the mean and the $2\sigma$ half-width of $x = 1/G$, respectively. The recovered conductance is therefore $\bar{G} = 1/\bar{x}$. The upper relative deviation of $G$ is obtained by propagating the lower limit $\bar{x} - 2\sigma_x$ through the reciprocal relation:

$$\delta_G^+ = \frac{1/(\bar{x}-2\sigma_x)-1/\bar{x}}{1/\bar{x}} = \frac{2\sigma_x}{\bar{x}-2\sigma_x} \tag{23}$$

Similarly, the lower relative deviation is

$$\delta_G^- = -\frac{2\sigma_x}{\bar{x}+2\sigma_x} \tag{24}$$

Since $2\sigma_x/\bar{x}$ is the 2RSD of $1/G$, the asymmetric relative uncertainty bounds can be written as

$$+\frac{2\mathrm{RSD}}{1-2\mathrm{RSD}} \text{ and } -\frac{2\mathrm{RSD}}{1+2\mathrm{RSD}}. \tag{25}$$

For the present Al–silica case, the analytical method gives $1/G_{\mathrm{Al/silica}} = 0.00952$ mm$^2$ K W$^{-1}$ with a 2RSD of 32%, corresponding to $2\sigma_x = 0.00305$ mm$^2$ K W$^{-1}$. After inversion, the recovered interfacial conductance is $G_{\mathrm{Al/silica}} = 105$ MW m$^{-2}$ K$^{-1}$, with asymmetric uncertainty bounds of $+47\%$ and $-24\%$, whereas a naive symmetric propagation in the original $G$ coordinate would give ±33%, underestimating the upper tail and overestimating the lower one. The Monte Carlo result [Fig. 5(b)] confirms this asymmetry, giving +56%/−24% around the same value: the lower bounds agree exactly, and both capture the pronounced positive skew. The larger MC upper bound reflects the difficulty of resolving the long upper tail with a finite sample, so the analytical result in the $1/G$ coordinate provides the more reliable upper bound. These results confirm that analytical uncertainty propagation should be performed in the natural coordinate $1/G$ and then converted back to $G$. This treatment restores approximate Gaussianity while preserving the physically correct asymmetric uncertainty structure of the recovered interfacial conductance. Table 2 summarizes this comparison for all recovered single parameters.

### *3.3 Geometric-mean conductivity and thermal effusivity of substrates*

Next, we test the natural coordinates for substrate-related properties. Two numerical cases are examined: Al–500 nm $SiO_2$–Si, where the substrate is isotropic and its heat capacity is weakly constrained; and Al–320 nm GaO–SiC, where the substrate is anisotropic and both in-plane and cross-plane conductivities are unknown. In both cases, synthetic experimental signals were generated from the forward model and then perturbed by added noise.

**Al–500 nm $SiO_2$–Si.** The known parameters are the Al transducer properties ($k_{\mathrm{Al}}$, $C_{\mathrm{Al}}$, $h_{\mathrm{Al}}$), the $SiO_2$ thickness $h_{\mathrm{SiO_2}}$, and the laser spot size $r_0$. The unknown fitting parameters are $G_{\mathrm{Al/SiO_2}}$, $k_{\mathrm{SiO_2}}$, $k_{\mathrm{Si}}$, $C_{\mathrm{SiO_2}}$, and $C_{\mathrm{Si}}$. Figure 6(a–e) shows the MC distributions obtained from the original variables. Among these, $G_{\mathrm{Al/SiO_2}}$ and $C_{\mathrm{Si}}$ exhibit evident right-skewed distributions, whereas $k_{\mathrm{SiO_2}}$, $k_{\mathrm{Si}}$, and $C_{\mathrm{SiO_2}}$ show approximately normal distributions.

To examine the transformed variables, the reciprocal of the interfacial thermal conductance and the substrate thermal effusivity were considered, while $k_{\mathrm{SiO_2}}$, $k_{\mathrm{Si}}$, and $C_{\mathrm{SiO_2}}$ were left untransformed. The resulting distributions are shown in Fig. 6(f–j). After transformation, the distributions of $k_{\mathrm{SiO_2}}$, $k_{\mathrm{Si}}$, and $C_{\mathrm{SiO_2}}$ remain essentially unchanged, further confirming that transforming selected parameters does not affect the distributions of the untransformed parameters. In contrast, both $1/G_{\mathrm{Al/SiO_2}}$ and $\sqrt{k_{\mathrm{Si}}C_{\mathrm{Si}}}$ become approximately normally distributed. The first result again confirms that the interfacial thermal resistance, rather than the interfacial thermal conductance, is the appropriate Gaussian variable. For the isotropic Si substrate, $k_{z,\mathrm{Si}} = k_{r,\mathrm{Si}} = k_{\mathrm{Si}}$, so the conductivity-related substrate variable $\sqrt{k_z k_r}$ reduces to $k_{\mathrm{Si}}$, whereas the effusivity-related variable $\sqrt{k_z C}$ becomes $\sqrt{k_{\mathrm{Si}} C_{\mathrm{Si}}}$. Therefore, this case supports the conclusion that the natural variables for substrate-related properties are $\sqrt{k_z k_r}$ and $\sqrt{k_z C}$.

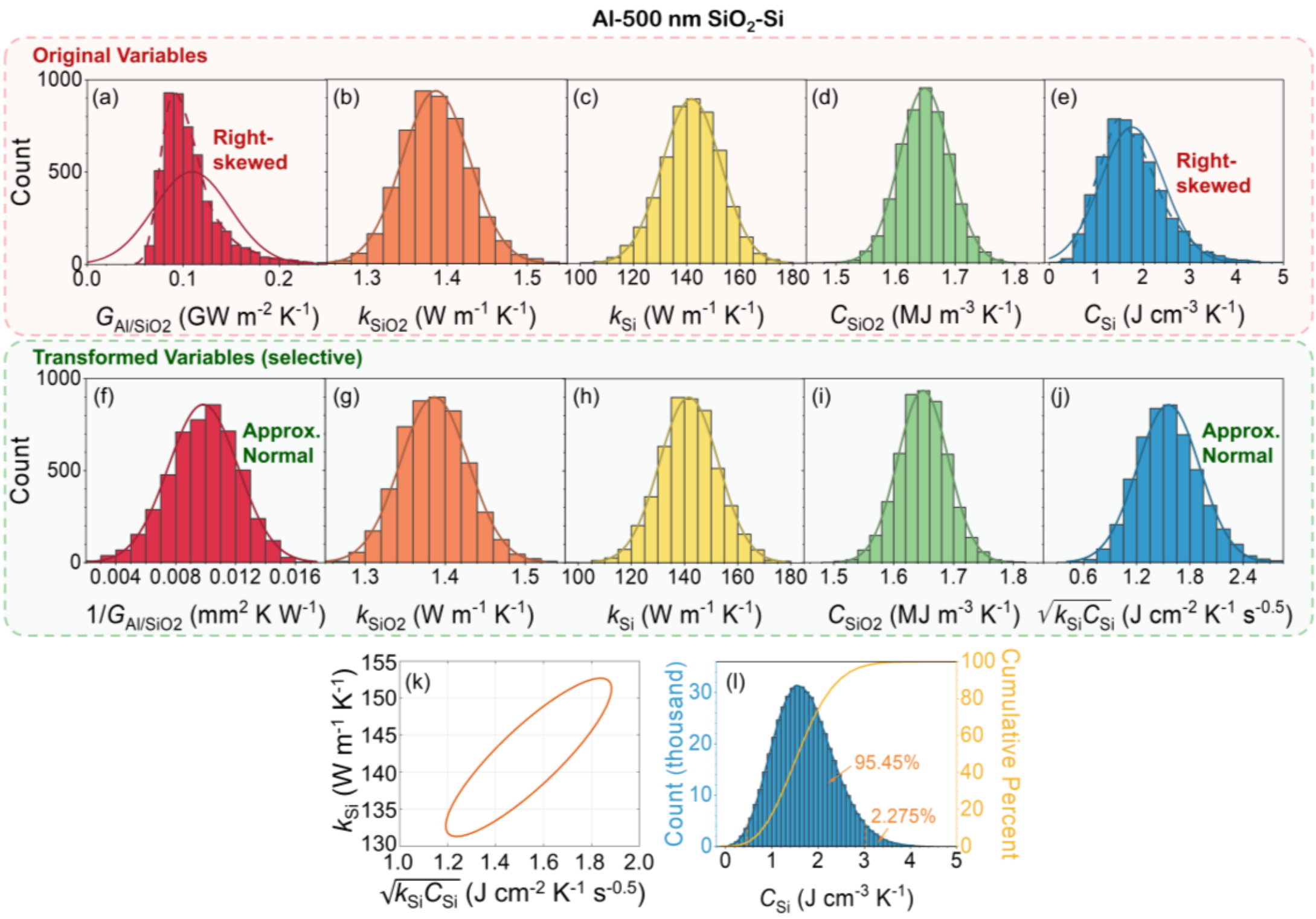

**FIG. 6.** Monte Carlo distributions for the Al–500 nm $SiO_2$–Si case. (a–e) Distributions of the original variables. Only (a) $G_{\mathrm{Al/SiO_2}}$ and (e) $C_{\mathrm{Si}}$ exhibit right-skewed distributions and therefore require transformation, whereas $k_{\mathrm{SiO_2}}$, $k_{\mathrm{Si}}$, and $C_{\mathrm{SiO_2}}$ are already approximately Gaussian in the original space. (f–j) Distributions of the five parameters after replacing $G_{\mathrm{Al/SiO_2}}$ with $1/G_{\mathrm{Al/SiO_2}}$ and $C_{\mathrm{Si}}$ with $\sqrt{k_{\mathrm{Si}} C_{\mathrm{Si}}}$, both of which show approximately normal distributions. (k) Joint normal distribution of $k_{\mathrm{Si}}$ and $\sqrt{k_{\mathrm{Si}} C_{\mathrm{Si}}}$ (ellipse: $2\sigma$ confidence contour). (l) Reconstructed distribution of $C_{\mathrm{Si}}$ obtained by sampling from the joint distribution in (k) and applying $C_{\mathrm{Si}} = (\sqrt{k_{\mathrm{Si}} C_{\mathrm{Si}}})^2 / k_{\mathrm{Si}}$.

Under this transformation, the analytical method can be applied directly to obtain the uncertainty of $1/G_{\mathrm{Al/SiO_2}}$, $k_{\mathrm{Si}}$, and $\sqrt{k_{\mathrm{Si}} C_{\mathrm{Si}}}$. The uncertainty of $G_{\mathrm{Al/SiO_2}}$ can then be recovered by the inverse transformation. $C_{\mathrm{Si}}$, however, is not one of the transformed variables; it must be reconstructed as $C_{\mathrm{Si}} = \left(\sqrt{k_{\mathrm{Si}}C_{\mathrm{Si}}}\right)^2/k_{\mathrm{Si}}$. Because this expression is a nonlinear combination of two correlated Gaussian variables, the resulting distribution of $C_{\mathrm{Si}}$ is inherently non-Gaussian and asymmetric; neither linear error propagation nor any closed-form formula yields reliable confidence bounds.

We therefore reconstruct the distribution of $C_{Si}$ by sampling from the joint Gaussian distribution of $k_{Si}$ and $\sqrt{k_{Si}C_{Si}}$ supplied by the analytical covariance matrix. The conductivity $k_{Si}$ and effusivity $e_{Si} = \sqrt{k_{Si}C_{Si}}$ are treated as a bivariate normal vector $\mathbf{y} = (k_{Si}, e_{Si})^T \sim \mathcal{N}(\boldsymbol{\mu}, \boldsymbol{\Sigma})$, with $\boldsymbol{\mu} = (\mu_k, \mu_e)^T$ and $\boldsymbol{\Sigma} = \begin{pmatrix} \sigma_k^2 & \mathrm{cov}(k_{Si}, e_{Si}) \\ \mathrm{cov}(k_{Si}, e_{Si}) & \sigma_e^2 \end{pmatrix} = \begin{pmatrix} \sigma_k^2 & \rho\sigma_k\sigma_e \\ \rho\sigma_k\sigma_e & \sigma_e^2 \end{pmatrix}$ taken from Eq. (16), where $\rho = \mathrm{cov}(k_{Si}, e_{Si})/(\sigma_k\sigma_e)$ is the Pearson correlation coefficient (PCC). Correlated samples are generated as $k_i = \mu_k + \sigma_k u_i$ and $e_i = \mu_e + \sigma_e(\rho u_i + \sqrt{1-\rho^2}\, v_i)$ with independent $u_i, v_i \sim \mathcal{N}(0,1)$, which reproduces $\boldsymbol{\mu}$ and $\boldsymbol{\Sigma}$ exactly, and each pair is mapped through $C_{Si,i} = e_i^2/k_i$. Over 500,000 draws of $(u_i, v_i)$, this yields the reconstructed distribution of $C_{Si}$ [Fig. 6(l)] in approximately 2 s, since each draw involves only random sampling and elementary algebra rather than any forward-model evaluation. This reconstruction framework is general: the distribution of any physical parameter can be reconstructed in the same way, provided it is a deterministic function of variables whose joint Gaussian distribution is supplied by the analytical method. We reuse it in the cases that follow.

**Al–320 nm GaO–SiC.** The known parameters include the Al transducer properties ($k_{\mathrm{Al}}$, $C_{\mathrm{Al}}$, $h_{\mathrm{Al}}$), the Al/GaO interfacial thermal conductance ($G_{\mathrm{Al/GaO}}$), the GaO film properties ($k_{\mathrm{GaO}}$, $C_{\mathrm{GaO}}$, $h_{\mathrm{GaO}}$), the SiC heat capacity $C_{\mathrm{SiC}}$, and the laser spot size $r_0$. The uncertainties of all thicknesses and $r_0$ were set to 2%, those of the volumetric heat capacities to 3%, and that of $k_{\mathrm{Al}}$ to 5%. The uncertainties of $G_{\mathrm{Al/GaO}}$ and $k_{\mathrm{GaO}}$ were taken from Ref. [28] as 17% and 6.3%, respectively. The unknown fitting parameters are $G_{\mathrm{GaO/SiC}}$, $k_{z,\mathrm{SiC}}$, and $k_{r,\mathrm{SiC}}$. Figure 7(a–c) shows the MC results for the original variables; all three exhibit clearly right-skewed distributions.

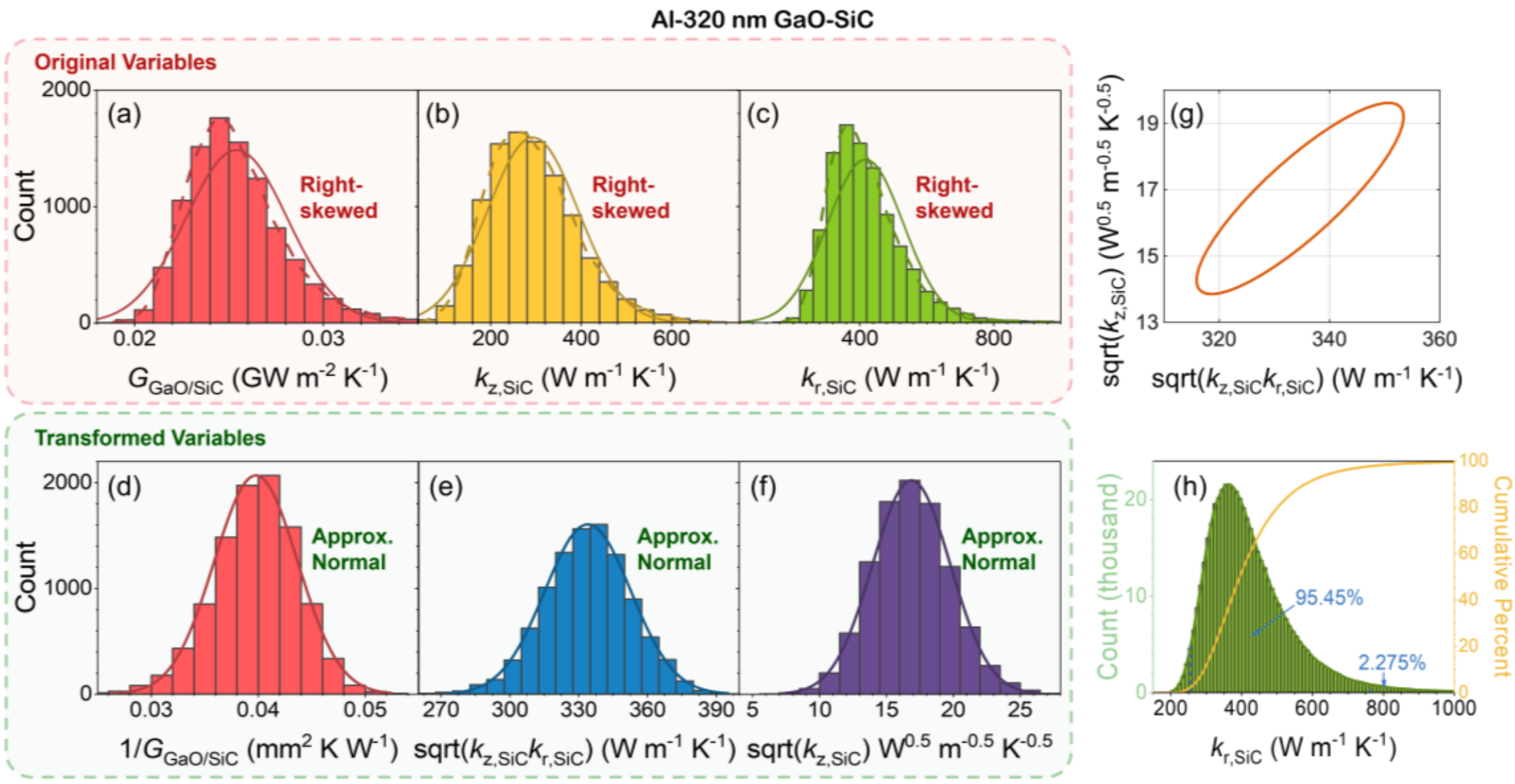

**FIG. 7.** Monte Carlo distributions for the Al–320 nm GaO–SiC case. (a–c) Original variables $G_{\mathrm{GaO/SiC}}$, $k_{z,\mathrm{SiC}}$, and $k_{r,\mathrm{SiC}}$ exhibit right-skewed distributions. (d–f) After transformation to $1/G_{\mathrm{GaO/SiC}}$, $\sqrt{k_{z,\mathrm{SiC}}}$, and $\sqrt{k_{z,\mathrm{SiC}}k_{r,\mathrm{SiC}}}$, all three become approximately normal. (g) Joint normal distribution of the two substrate-related transformed variables (ellipse: $2\sigma$ confidence contour). (h) Reconstructed distribution of $k_{r,\mathrm{SiC}}$ obtained by sampling from the joint distribution and applying $k_{r,\mathrm{SiC}} = \left(\sqrt{k_{z,\mathrm{SiC}}k_{r,\mathrm{SiC}}}\right)^2/k_{z,\mathrm{SiC}}$.

The MC distributions for the transformed variables of $1/G_{\mathrm{GaO/SiC}}$, $\sqrt{k_{z,\mathrm{SiC}}k_{r,\mathrm{SiC}}}$, and $\sqrt{k_{z,\mathrm{SiC}}}$ are shown in Fig. 7(d–f). After transformation, all three distributions become approximately normal. Here, $\sqrt{k_{z,\mathrm{SiC}}k_{r,\mathrm{SiC}}}$ corresponds to the geometric-mean thermal conductivity of the anisotropic substrate, while $\sqrt{k_{z,\mathrm{SiC}}}$ is proportional to $\sqrt{k_{z,\mathrm{SiC}}C_{\mathrm{SiC}}}$ because $C_{\mathrm{SiC}}$ is treated as a known parameter. These results provide direct evidence that the natural variables associated with substrate properties are $\sqrt{k_z k_r}$ and $\sqrt{k_z C}$, rather than $k_z$ and $k_r$ separately.

Based on these transformed variables, the covariance-based analytical method can directly determine the uncertainties of $1/G_{\mathrm{GaO/SiC}}$, $\sqrt{k_{z,\mathrm{SiC}}k_{r,\mathrm{SiC}}}$, and $\sqrt{k_{z,\mathrm{SiC}}}$. The uncertainties of $G_{\mathrm{GaO/SiC}}$ and $k_{z,\mathrm{SiC}}$ can then be obtained by simple algebraic inversion. Recovering $k_{r,SiC}$, however, requires the joint distribution of $\sqrt{k_{z,SiC}k_{r,SiC}}$ and $\sqrt{k_{z,SiC}}$ [Fig. 7(g)], which the analytical covariance matrix provides directly. Applying the same analytical-covariance reconstruction framework used for $C_{Si}$ above, with the inverse map $k_{r,SiC} = (\sqrt{k_{z,SiC}k_{r,SiC}})^2/(\sqrt{k_{z,SiC}})^2$, gives the reconstructed distribution in Fig. 7(h), from which the central 95.45% bounds are read off.

The recovery of $k_{z,\mathrm{SiC}}$ from its natural coordinate illustrates another elementary inversion, in which the physical parameter is obtained by squaring an approximately Gaussian variable. Let $s$ and $\sigma_s$ denote the mean and standard uncertainty of $\sqrt{k_z}$, respectively. The recovered value is $k_{z,0} = s^2$, and the symmetric 95.45% interval in the transformed coordinate is $s \pm 2\sigma_s$. Squaring these limits gives the relative deviations in the original $k_z$ coordinate:

$$\delta_{k_z}^{+} = \frac{(s+2\sigma_s)^2 - s^2}{s^2} = \left(1+\frac{2\sigma_s}{s}\right)^2 - 1 \tag{26}$$

and

$$\delta_{k_z}^{-} = \left(1-\frac{2\sigma_s}{s}\right)^2 - 1 \tag{27}$$

Here, $2\sigma_s/s$ is the 2RSD of $\sqrt{k_z}$. For the Al–320 nm GaO–SiC case, $\sqrt{k_{z,\mathrm{SiC}}}$ has a 2RSD of 34.5%. Therefore, the recovered value $k_{z,\mathrm{SiC}} = 280\ \mathrm{W\,m^{-1}\,K^{-1}}$ corresponds to asymmetric uncertainty bounds of +81%/-57%. The directly sampled MC distribution [Fig. 7(b)] gives +87%/-55% around the same recovered value, confirming the validity of the quadratic inversion, as summarized in Table 2.

Taken together, these two cases show that substrate-related parameters are more naturally described in terms of $\sqrt{k_z k_r}$ and $\sqrt{k_z C}$ than in terms of the original variables $k_z$, $k_r$, and $C$. This finding explains the skewed Monte Carlo distributions commonly observed in the original parameter space and provides the appropriate coordinates for fast analytical uncertainty propagation.

*3.4 Diffusivity, areal heat capacity, and diffusion time for thin films*

We finally examine the natural coordinates for thin-film-related properties using three cases of increasing complexity: Al–2 μm $SiO_2$–Si and Al–1.59 μm Si–1.03 μm $SiO_2$–Si, which are numerical cases with synthetic experimental signals, and the experimental case of Al–1.08 μm GaN–0.46 μm AlGaN–Si introduced in Section 3.1.

**Al–2 μm $SiO_2$–Si.** The known parameters include the Al transducer properties ($k_{\mathrm{Al}}$, $C_{\mathrm{Al}}$, $h_{\mathrm{Al}}$), the $SiO_2$ heat capacity and thickness ($C_{\mathrm{SiO_2}}$, $h_{\mathrm{SiO_2}}$), the deep interface conductance ($G_{\mathrm{SiO_2/Si}}$), the Si substrate properties ($k_{\mathrm{Si}}$, $C_{\mathrm{Si}}$), and the laser spot size $r_0$. The uncertainties of the thermal conductivities were set to 5%, those of the volumetric heat capacities to 3%, and those of the thicknesses and $r_0$ to 2%. The interfacial thermal conductance $G_{\mathrm{SiO_2/Si}}$ was fixed at 200 MW $m^{-2}$ $K^{-1}$; because its sensitivity is extremely weak, its uncertainty was set to 50%. The unknown fitting parameters are $G_{\mathrm{Al/SiO_2}}$, $k_{z,\mathrm{SiO_2}}$, and $k_{r,\mathrm{SiO_2}}$. Figure 8(a–c) shows the distributions of the original variables. Among these parameters, $G_{\mathrm{Al/SiO_2}}$ is clearly right-skewed, whereas $k_{r,\mathrm{SiO_2}}$ remains approximately normal even though its uncertainty is relatively large (RSD≈74%). Because $C_{\mathrm{SiO_2}}$ is treated as known in this case, the normality of $k_{r,\mathrm{SiO_2}}$ directly supports the conclusion that the thin-film variable $k_r/C$ is approximately Gaussian: $k_{r,\mathrm{SiO_2}}$ and $k_{r,\mathrm{SiO_2}}/C_{\mathrm{SiO_2}}$ differ only by the known constant $C_{\mathrm{SiO_2}}$. After replacing $G_{\mathrm{Al/SiO_2}}$ with its reciprocal, the transformed distribution [Fig. 8(d)] becomes approximately normal, again confirming that the interfacial thermal resistance $1/G$ is the appropriate Gaussian variable.

**Al–1.08 μm GaN–0.46 μm AlGaN–Si.** We revisit the experimental system introduced in Section 3.1, now focusing specifically on the thin-film variables. In the original parameter space, $k_{z,\mathrm{GaN}}$ and $k_{\mathrm{AlGaN}}$ show clearly right-skewed distributions [Fig. 4(g, h)]. When these two parameters are replaced by their reciprocals $1/k_{z,\mathrm{GaN}}$ and $1/k_{\mathrm{AlGaN}}$, and the fitting is repeated with the five-parameter set, the Monte Carlo results [Fig. 9(a–e)] indicate that all transformed variables become approximately normal. In this case, the reciprocal variables reflect the thin-film quantity $Ch^2/k_z$, because the corresponding volumetric heat capacities and thicknesses of GaN and AlGaN are known.

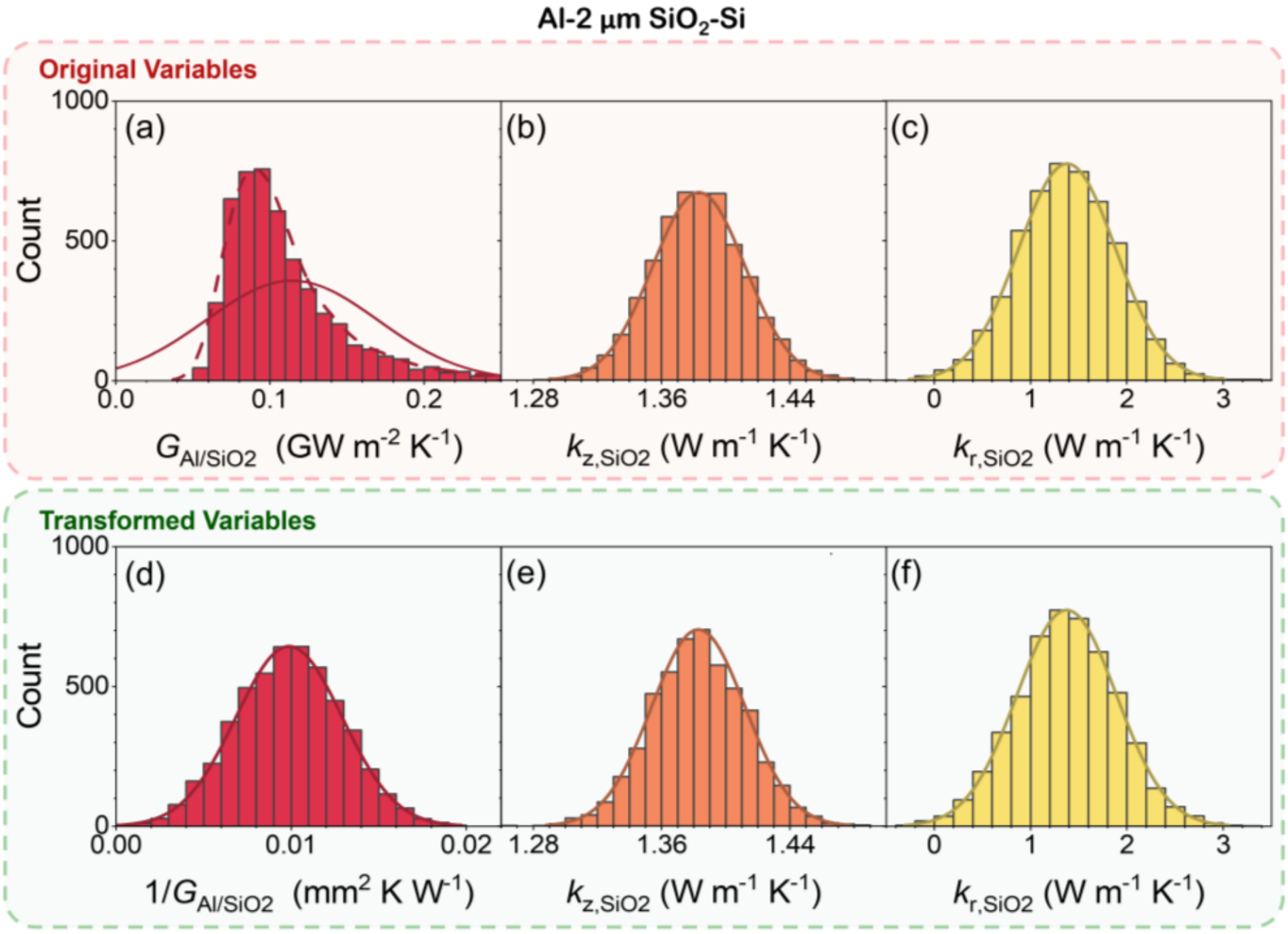

**FIG. 8.** Monte Carlo distributions of the fitted parameters for the Al–2 μm $SiO_2$–Si case. (a–c) Distributions obtained for $G_{\mathrm{Al/SiO_2}}$, $k_{z,\mathrm{SiO_2}}$, and $k_{r,\mathrm{SiO_2}}$ in the original parameter space. (d–f) Distributions of the three parameters after replacing $G_{\mathrm{Al/SiO_2}}$ with $1/G_{\mathrm{Al/SiO_2}}$.

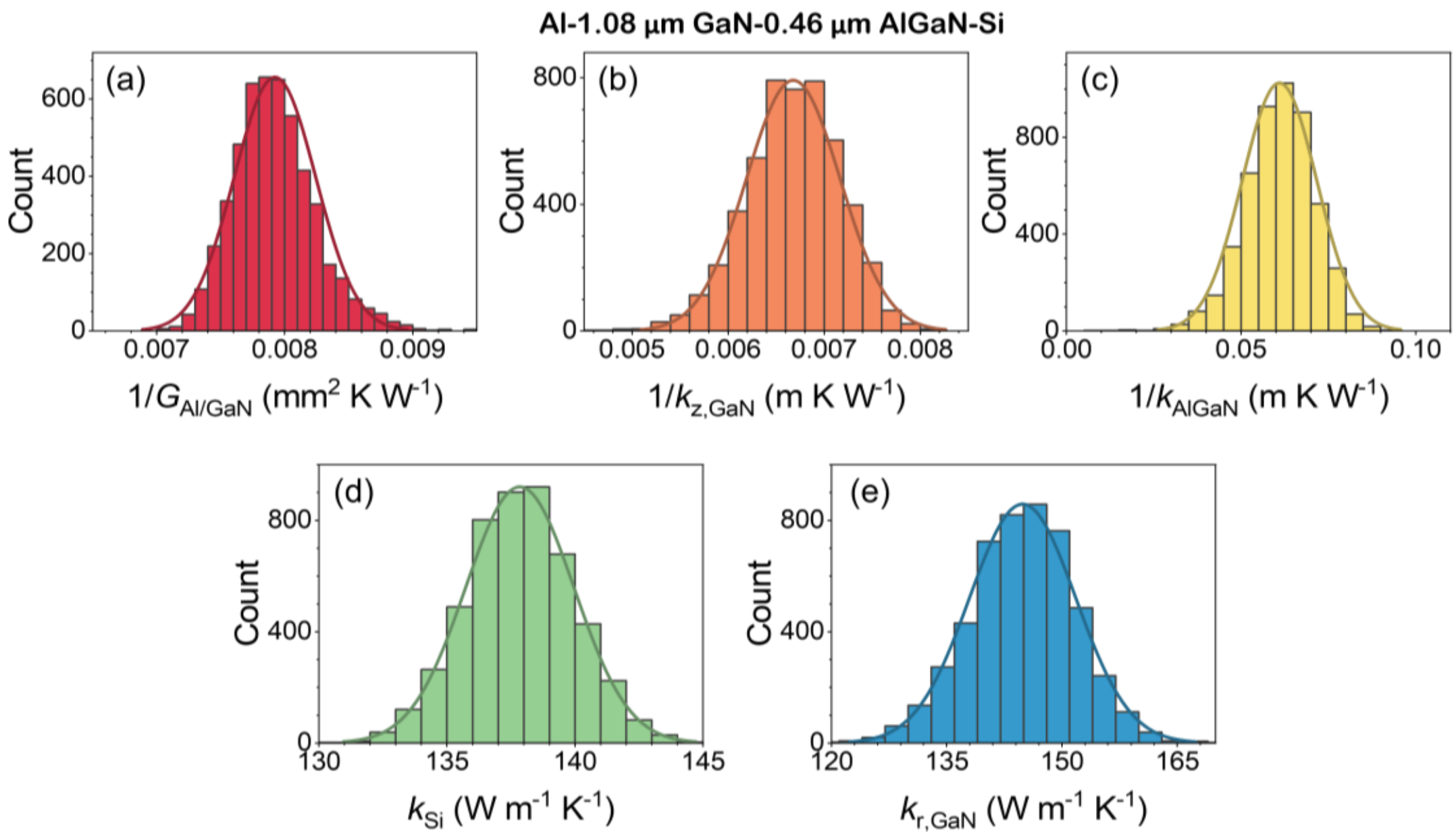

**FIG. 9.** Monte Carlo distributions of the transformed fitted parameters for the Al–1.08 μm GaN–0.46 μm AlGaN–Si case. After replacing $G_{\mathrm{Al/GaN}}$, $k_{z,\mathrm{GaN}}$, and $k_{\mathrm{AlGaN}}$ by their reciprocal forms, all five fitted variables become approximately normally distributed.

**Al−1.59 μm Si−1.03 μm $SiO_2$−Si.** This is the most complex case, with six simultaneously fitted unknowns: $G_{\mathrm{Al/Si}}$, $k_{z,\mathrm{Si(f)}}$, $k_{r,\mathrm{Si(f)}}$, $C_{\mathrm{Si(f)}}$, $k_{\mathrm{SiO_2}}$, and $C_{\mathrm{SiO_2}}$. The known parameters are $k_{\mathrm{Al}}$, $C_{\mathrm{Al}}$, $h_{\mathrm{Al}}$, $h_{\mathrm{Si(f)}}$, $h_{\mathrm{SiO_2}}$, $k_{\mathrm{Si(sub)}}$, $C_{\mathrm{Si(sub)}}$, $G_{\mathrm{Si/SiO_2}}$, and $r_0$, with values taken from Ref. [29]. The distributions of the six unknown parameters are shown in Fig. 10(a–f).

To further examine the thin-film variables, $G_{\mathrm{Al/Si}}$, $k_{z,\mathrm{Si(f)}}$, and $k_{\mathrm{SiO_2}}$ were replaced by their reciprocals. The resulting distributions [Fig. 10(g–i)] show that both $1/G_{\mathrm{Al/Si}}$ and $1/k_{\mathrm{SiO_2}}$ remain approximately normal. By contrast, $1/k_{z,\mathrm{Si(f)}}$ does not become normally distributed. This does not contradict the conclusion that $C/k_z$ is Gaussian; it reflects the fact that $k_z$, $k_r$, and $C$ are fitted simultaneously and therefore influence one another statistically. When the variables are instead expressed as $C_{\mathrm{Si(f)}}/k_{z,\mathrm{Si(f)}}$ and $k_{r,\mathrm{Si(f)}}/C_{\mathrm{Si(f)}}$, the resulting distributions [Fig. 10(j–k)] are both approximately normal, directly validating the proposed thin-film classification: $C/k_z$ (cross-plane diffusion time) and $k_r/C$ (in-plane thermal diffusivity).

To recover the uncertainties of $k_{z,\mathrm{Si(f)}}$ and $k_{r,\mathrm{Si(f)}}$, the joint normal distribution of $C_{\mathrm{Si(f)}}/k_{z,\mathrm{Si(f)}}$, $k_{r,\mathrm{Si(f)}}/C_{\mathrm{Si(f)}}$, and $C_{\mathrm{Si(f)}}$ must be considered, as shown in Fig. 10(l). This joint distribution is obtained directly from the covariance matrix in the analytical method. Sampling 500,000 random triplets from this joint distribution and calculating the corresponding values of $k_{z,\mathrm{Si(f)}}$ and $k_{r,\mathrm{Si(f)}}$ yields the reconstructed distributions shown in Fig. 10(m, n). Because this procedure involves only simple multiplication and division, the additional computational cost is negligible. The lower and upper bounds of $k_{z,\mathrm{Si(f)}}$ and $k_{r,\mathrm{Si(f)}}$ were then determined from the central 95.45% confidence interval.

Taken together, these three cases show that thin-film-related parameters are more naturally described in terms of $k_r/C$, $hC$, and $Ch^2/k_z$ than in terms of the original variables $k_r$, $C$, and $k_z$ separately. This classification explains why some thin-film parameters appear skewed in the original parameter space and provides the appropriate transformed variables for fast and reliable analytical uncertainty propagation.

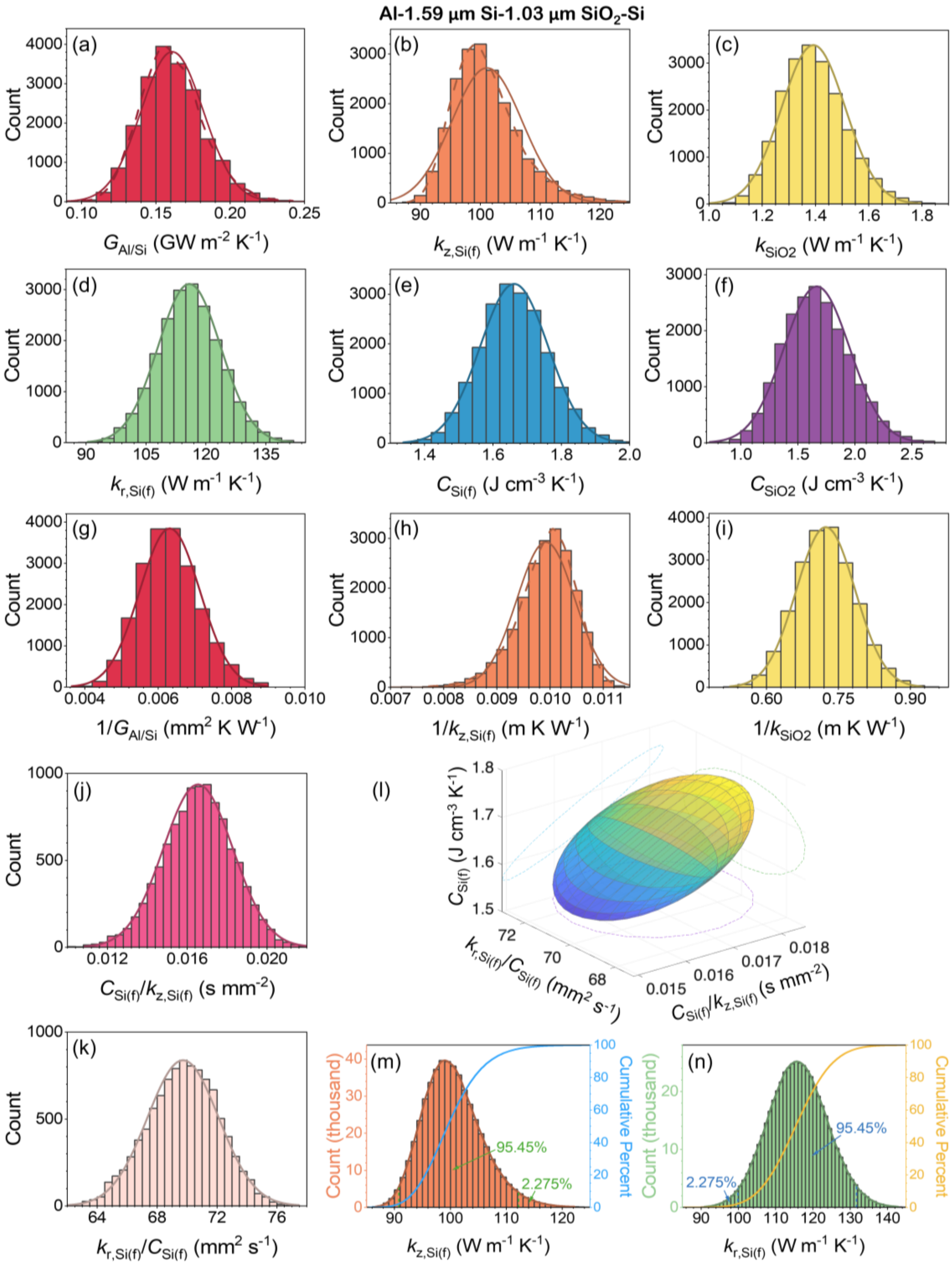

**FIG. 10.** Monte Carlo and reconstructed distributions for the Al–1.59 μm Si–1.03 μm $SiO_2$–Si case. (a–f) Distributions of the fitted parameters in the original parameter space. (g–i) Distributions after reciprocal transformation of $G_{\mathrm{Al/Si}}$, $k_{z,\mathrm{Si(f)}}$, and $k_{\mathrm{SiO_2}}$. (j–k) Distributions of the transformed thin-film variables $C_{\mathrm{Si(f)}}/k_{z,\mathrm{Si(f)}}$ and $k_{r,\mathrm{Si(f)}}/C_{\mathrm{Si(f)}}$. (l) Joint normal distribution of $C_{\mathrm{Si(f)}}/k_{z,\mathrm{Si(f)}}$, $k_{r,\mathrm{Si(f)}}/C_{\mathrm{Si(f)}}$, and $C_{\mathrm{Si(f)}}$. (m – n) Reconstructed distributions of $k_{z,\mathrm{Si(f)}}$ and $k_{r,\mathrm{Si(f)}}$ from the joint distribution.

*3.5 Summary of natural coordinates, practical workflow, and limitations*

Sections 3.2 – 3.4 have established, through six representative FDTR multilayer systems, that thermoreflectance inverse problems possess an intrinsic statistical geometry rooted in the multilayer heat diffusion equation. The approximately Gaussian variables are not necessarily the original physical parameters themselves, but specific transformed coordinates: the reciprocal conductance $1/G$ for interfaces; the geometric-mean thermal conductivity $\sqrt{k_z k_r}$ and thermal effusivity $\sqrt{k_z C}$ for substrates; and the in-plane thermal diffusivity $k_r/C$, the areal heat capacity $hC$, and the cross-plane diffusion time $Ch^2/k_z$ for thin films. Across the six representative cases, the marginal distributions of these natural coordinates are approximately Gaussian, providing the statistical basis for covariance-based analytical uncertainty propagation in the transformed space.

Because marginal Gaussianity alone does not guarantee multivariate Gaussianity, the joint distributions of the natural coordinates were further assessed using the squared Mahalanobis distances [30] of the normalized-MC samples. The results show approximately multivariate-Gaussian behavior across the six representative cases, with detailed $\chi^2$ Q–Q and cumulative-distribution comparisons provided in Fig. S2 of the Supplementary Information.

To quantify the fidelity of the transformed analytical method relative to the normalized Monte Carlo benchmark, Table 1 summarizes the mean values and relative uncertainties for all six representative FDTR cases. Across the full range of parameter categories—interfacial, substrate, and thin-film—the analytical results agree closely with the Monte Carlo benchmarks. Typical discrepancies in the inferred relative uncertainties remain within a few percentage points, even for parameters with relatively large uncertainties. For example, the relative uncertainty of $1/G_{\mathrm{Al/SiO_2}}$ in the Al–2 μm $SiO_2$–Si case is 60.7% from Monte Carlo and 63.4% from the analytical method. This agreement confirms that, once the appropriate natural coordinates are adopted, the local linearization underlying covariance-based analytical propagation becomes sufficiently accurate for the thermoreflectance inverse problems, despite the marked non-Gaussian behavior observed in the original parameter space.

**Table 1.** Comparison of the mean values and relative uncertainties obtained from Monte Carlo simulations and the analytical method for all representative cases. The relative uncertainty is reported as 2RSD.

| Case | Parameter | MC mean | MC 2RSD (%) | Analytical mean | Analytical 2RSD (%) |
|---|---|---|---|---|---|
| Al–silica | $k_{\mathrm{Al}}$ | 123 | 4.5 | 123 | 6.1 |
| | $1/G_{\mathrm{Al/silica}}$ | 0.00945 | 34 | 0.00952 | 32 |
| | $k_{\mathrm{silica}}$ | 1.38 | 3.8 | 1.38 | 4.5 |
| | $C_{\mathrm{silica}}$ | 1.65 | 4.1 | 1.65 | 4.4 |
| Al–500 nm $SiO_2$–Si | $1/G_{\mathrm{Al/SiO2}}$ | 0.0099 | 47 | 0.010 | 47 |
| | $k_{\mathrm{SiO2}}$ | 1.39 | 6.1 | 1.38 | 6.0 |
| | $k_{\mathrm{Si}}$ | 142 | 15.6 | 142 | 15.2 |
| | $C_{\mathrm{SiO2}}$ | 1.65 | 5.1 | 1.65 | 5.0 |
| | $\sqrt{k_{\mathrm{Si}} C_{\mathrm{Si}}}$ | 1.56 | 45.2 | 1.54 | 45.0 |
| Al–320 nm GaO–SiC | $1/G_{\mathrm{GaO/SiC}}$ | 0.0398 | 19.6 | 0.0400 | 19.0 |
| | $\sqrt{k_{\mathrm{z,SiC}} k_{\mathrm{r,SiC}}}$ | 334.0 | 11.3 | 334.7 | 11.2 |
| | $\sqrt{k_{\mathrm{z,SiC}}}$ | 16.87 | 34.4 | 16.73 | 34.5 |
| Al–2 μm $SiO_2$–Si | $1/G_{\mathrm{Al/SiO2}}$ | 0.0099 | 60.7 | 0.0100 | 63.4 |
| | $k_{\mathrm{z,SiO2}}$ | 1.38 | 4.2 | 1.38 | 4.2 |
| | $k_{\mathrm{r,SiO2}}$ | 1.38 | 74.3 | 1.38 | 80.7 |
| Al–1.08 μm GaN–0.46 μm AlGaN–Si | $1/G_{\mathrm{Al/GaN}}$ | 0.0079 | 8.0 | 0.0079 | 7.6 |
| | $1/k_{\mathrm{z,GaN}}$ | 0.0067 | 14.9 | 0.0067 | 14.2 |
| | $1/k_{\mathrm{AlGaN}}$ | 0.0610 | 35.1 | 0.0625 | 30.4 |
| | $k_{\mathrm{Si}}$ | 137.8 | 3.1 | 138.0 | 3.8 |
| | $k_{\mathrm{r,GaN}}$ | 144.9 | 9.6 | 145.0 | 10.6 |
| Al–1.59 μm Si–1.03 μm $SiO_2$–Si | $1/G_{\mathrm{Al/Si}}$ | 0.0063 | 26.0 | 0.0063 | 25.2 |
| | $C_{\mathrm{Si(f)}}/k_{\mathrm{z,Si(f)}}$ | 0.0165 | 21.2 | 0.0166 | 20.4 |
| | $k_{\mathrm{r,Si(f)}}/C_{\mathrm{Si(f)}}$ | 69.72 | 6.8 | 69.88 | 7.8 |
| | $C_{\mathrm{Si(f)}}$ | 1.66 | 12.0 | 1.66 | 11.7 |
| | $1/k_{\mathrm{SiO2}}$ | 0.7236 | 17.0 | 0.7246 | 16.9 |
| | $C_{\mathrm{SiO2}}$ | 1.67 | 34.6 | 1.65 | 34.3 |

Note: $k$ and $\sqrt{k_z k_r}$ are in W m$^{-1}$ K$^{-1}$; $1/G$ is in mm$^2$ K W$^{-1}$; $C$ is in J cm$^{-3}$ K$^{-1}$; $\sqrt{kC}$ is in J cm$^{-2}$ K$^{-1}$ s$^{-0.5}$; $C/k_z$ is in s mm$^{-2}$; and $k_r/C$ is in mm$^2$ s$^{-1}$.

The consistency observed in Table 1 has direct practical implications. Once the natural coordinates are identified, the analytical method provides a computationally efficient alternative to full Monte Carlo analysis without sacrificing statistical fidelity. For a typical multilayer thermoreflectance fit involving three to six unknowns and multiple datasets, the analytical route reduces the uncertainty-evaluation time from repeated nonlinear optimizations to a single evaluation of the Jacobian and covariance matrices (approximately 0.6 s). The complete uncertainty evaluation requires a computational time comparable to that of a single MC refit. In contrast, full MC uncertainty propagation requires thousands of such nonlinear refits, resulting in orders-of-magnitude greater computational cost. The substantial efficiency gain arises because the transformation identifies the coordinates in which the inverse problem is locally closer to linear and the fitted distributions are closer to Gaussian.

Equations (23) and (24), together with the joint-sampling reconstruction described above, make the asymmetric uncertainty of every recovered physical parameter available analytically at negligible additional cost. Table 2 compares these analytical bounds with the central 95.45% intervals read directly from the full Monte Carlo distributions for all recovered single parameters, spanning the interfacial, substrate, and thin-film categories. In every case, the analytical method recovers the correct direction of the asymmetry, typically with a larger positive than negative deviation. The magnitudes also agree closely: averaged over the recovered parameters, the lower bounds differ by only a few percentage points, while the upper bounds show larger but still acceptable deviations. The largest discrepancies occur for strongly right-skewed interfacial conductances, such as $G_{\mathrm{Al/silica}}$ and the weakly sensitive $G_{\mathrm{Al/SiO2}}$ of the 2-μm-film case, whose extreme upper tails are governed by residual non-Gaussianity and are therefore the most sensitive to finite-sample noise. Even in these cases, however, the transformed analytical interval is far more faithful than a symmetric ±2σ interval evaluated directly in the original conductance coordinate, which would underestimate the positive tail and overestimate the negative one.

**Table 2.** Asymmetric uncertainty of the recovered single parameters: central 95.45% confidence intervals (+upper/−lower, relative to the recovered value) obtained directly from the full Monte Carlo distributions and from the analytical inversion of the natural coordinates [Eqs. (23-24) and (26-27) for the reciprocal and quadratic maps; joint-distribution sampling for parameters recovered from coupled coordinates].

| Case | Parameter | Inverse map | Recovered value | MC 95.45% CI | Analytical 95.45% CI |
|---|---|---|---|---|---|
| Al–silica | $G_{\mathrm{Al/silica}}$ | $1/G$ | 105 MW m$^{-2}$ K$^{-1}$ | +56% / −24% | +47% / −24% |
| Al–500 nm $SiO_2$–Si | $G_{\mathrm{Al/SiO2}}$ | $1/G$ | 100 MW m$^{-2}$ K$^{-1}$ | +101% / −30% | +89% / −32% |
| | $C_{\mathrm{Si}}$ | joint | 1.67 J cm$^{-3}$K$^{-1}$ | +101% / −59% | +87% / −65% |
| Al–320 nm GaO–SiC | $G_{\mathrm{GaO/SiC}}$ | $1/G$ | 25 MW m$^{-2}$ K$^{-1}$ | +28% / −16% | +23% / −16% |
| | $k_{\mathrm{z,SiC}}$ | $\sqrt{k_z}$ | 280 W m$^{-1}$ K$^{-1}$ | +87% / −55% | +81% / −57% |
| | $k_{\mathrm{r,SiC}}$ | joint | 400 W m$^{-1}$ K$^{-1}$ | +74% / −37% | +92% / −35% |
| Al–2 μm $SiO_2$–Si | $G_{\mathrm{Al/SiO2}}$ | $1/G$ | 100 MW m$^{-2}$ K$^{-1}$ | +155% / −37% | +173% / −39% |
| Al–1.08 μm GaN–0.46 μm AlGaN–Si | $G_{\mathrm{Al/GaN}}$ | $1/G$ | 127 MW m$^{-2}$ K$^{-1}$ | +8% / −7% | +7% / −9% |
| | $k_{\mathrm{z,GaN}}$ | $1/k_z$ | 149 W m$^{-1}$ K$^{-1}$ | +19% / −12% | +17% / −12% |
| | $k_{\mathrm{AlGaN}}$ | $1/k$ | 16 W m$^{-1}$ K$^{-1}$ | +54% / −22% | +44% / −23% |
| Al–1.59 μm Si–1.03 μm $SiO_2$–Si | $G_{\mathrm{Al/Si}}$ | $1/G$ | 160 MW m$^{-2}$ K$^{-1}$ | +29% / −23% | +34% / −20% |
| | $k_{\mathrm{z,Si(f)}}$ | joint | 100 W m$^{-1}$ K$^{-1}$ | +12% / −8% | +13% / −9% |
| | $k_{\mathrm{r,Si(f)}}$ | joint | 116 W m$^{-1}$ K$^{-1}$ | +14% / −14% | +14% / −14% |
| | $k_{\mathrm{SiO2}}$ | $1/k$ | 1.38 W m$^{-1}$ K$^{-1}$ | +20% / −15% | +20% / −15% |

Beyond the FDTR cases examined above, additional MC validations using SPS, TDTR, and SSTR further show that the same natural coordinates remain approximately Gaussian and that the analytical uncertainty estimates remain consistent with the MC results. Detailed results are provided in Section S3 of the Supplementary Information.

To operationalize these findings, we propose the following workflow for practitioners, as illustrated in Fig. 11.

First, identify the parameter category. The unknown physical parameter of interest $P$ should be classified according to whether it belongs to an interface, a semi-infinite substrate, or a finite thin film. This classification follows directly from the layer geometry of the sample under investigation.

Second, select the corresponding natural coordinate $X$. For interfacial transport, the appropriate coordinate is the reciprocal conductance $1/G$, i.e., the interfacial thermal resistance. For substrate properties, the natural coordinates are the geometric-mean thermal conductivity $\sqrt{k_z k_r}$ and the thermal effusivity $\sqrt{k_z C}$. For isotropic substrates, $\sqrt{k_z k_r}$ reduces to $k$. For thin-film properties, the natural coordinates are the in-plane thermal diffusivity $k_r/C$, the areal heat capacity $hC$, and the cross-plane diffusion time $Ch^2/k_z$, or equivalently $C/k_z$ when the thickness is independently known. In the most general setting, $m$ correlated physical parameters $\mathbf{P} = (P_1, \ldots, P_m)^{\mathrm{T}}$ are recovered from $n$ correlated natural coordinates $\mathbf{X} = (X_1, \ldots, X_n)^{\mathrm{T}} \in \mathbb{R}^n$ via $\mathbf{P} = f(\mathbf{X})$, where $f = (f_1, \ldots, f_m)^{\mathrm{T}}$ is a vector of $m$ known functions, one $f_j$ per $P_j$. Since each $X_i$ is approximately Gaussian by construction, $\mathbf{X} \sim \mathcal{N}(\boldsymbol{\mu}, \boldsymbol{\Sigma})$, where the mean vector $\boldsymbol{\mu} \in \mathbb{R}^n$ and covariance matrix $\boldsymbol{\Sigma} \in \mathbb{R}^{n\times n}$ are obtained from Eqs. (13)–(15) at the cost of a single Jacobian evaluation at the optimal fit.

Third, recover the uncertainty of $\mathbf{P}$ from natural coordinate $\mathbf{X}$ by analytical-covariance Monte Carlo sampling. Samples of $\mathbf{X}$ are drawn as $\mathbf{X}_i = \boldsymbol{\mu} + L\mathbf{z}_i$, where $L \in \mathbb{R}^{n\times n}$ satisfies $\boldsymbol{\Sigma} = LL^{\mathrm{T}}$ and $\mathbf{z}_i \sim \mathcal{N}(\mathbf{0}, I_n)$ are independent. A single draw sequence $\{\mathbf{z}_i\}$ yields every $f_j(\mathbf{X}_i)$ from the same realization, automatically preserving all cross-correlations among the recovered parameters; the bounds of each $P_j$ are read from the central 95.45% interval of $\{f_j(\mathbf{X}_i)\}$. When a particular $f_j$ depends on a single coordinate, the inverse relation can alternatively be applied in closed form to that coordinate's symmetric interval: Eqs. (23)–(24) for a reciprocal map (e.g., $G = 1/(1/G)$) or Eqs. (26)–(27) for a quadratic map (e.g., $k_z = (\sqrt{k_z})^2$). Because this reconstruction involves only arithmetic operations on the already-computed covariance matrix, its cost is negligible compared with full Monte Carlo refitting.

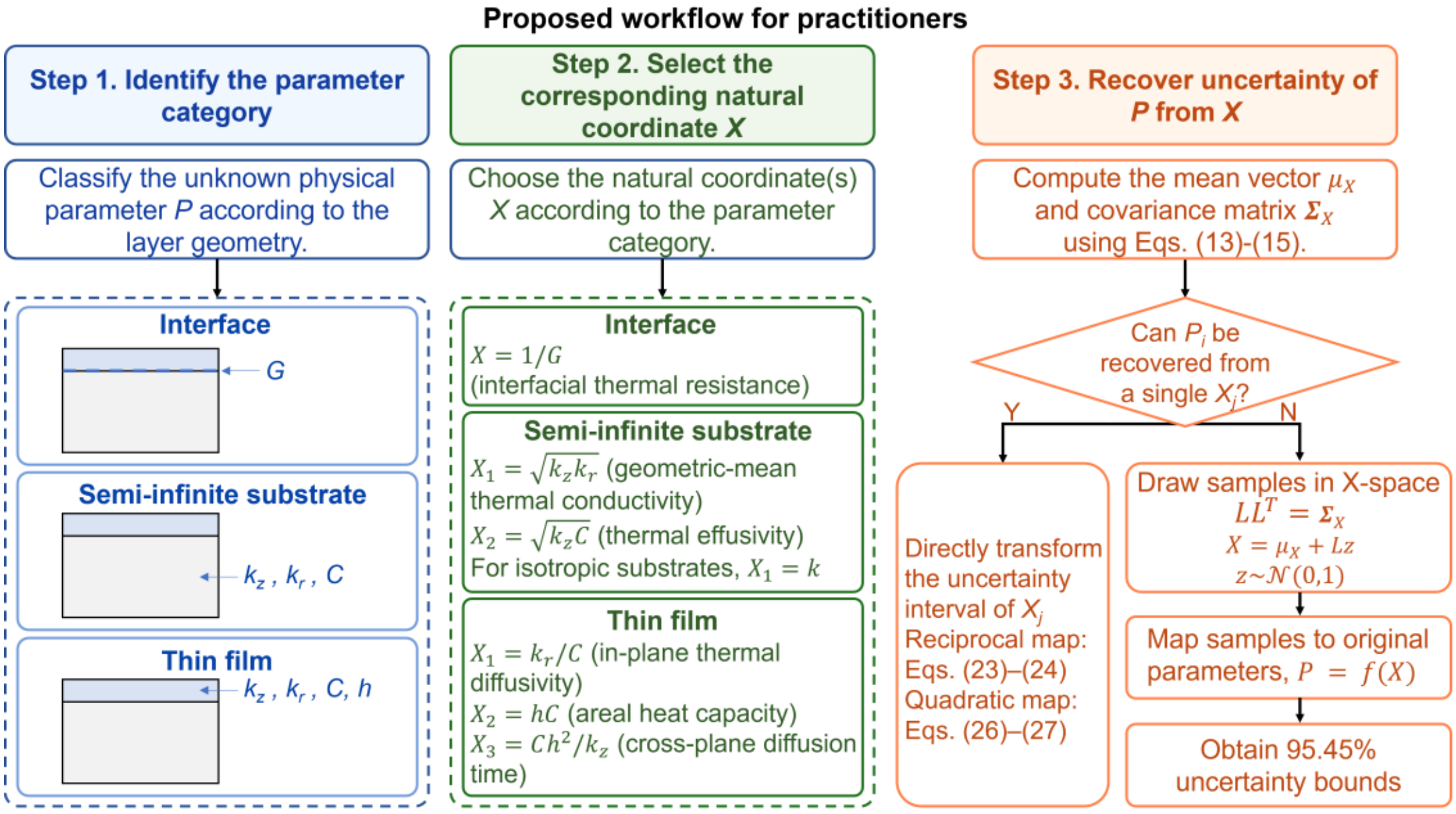

**FIG. 11.** Workflow for practitioners to select natural coordinates and recover parameter uncertainties.

The practical relevance of the proposed framework can be illustrated by several engineering scenarios. In AlGaN/GaN power-electronic heterostructures, the thermal conductivities of the constituent layers and interfacial thermal resistances are important inputs for device-level thermal simulations [31], and their uncertainties therefore affect the confidence of predicted operating temperatures. For anisotropic wide-bandgap semiconductors such as β-$Ga_2O_3$, directional thermal conductivity is an important consideration in high-power thermal management [32]; the substrate natural coordinates identified here provide a physically motivated basis for quantifying the coupled uncertainty of in-plane and cross-plane thermal transport properties. In three-dimensional heterogeneous integration, FDTR has been used to characterize the thermal resistance of fusion-bonded Si/$SiO_2$ stacks, where buried dielectric layers and interfaces can form important thermal bottlenecks [33]. The interface and thin-film natural coordinates developed here can provide a computationally efficient route for quantifying uncertainty in such fitted thermal properties. These examples illustrate how the proposed framework can support thermal characterization for device modeling, material and process comparison, and multilayer thermal-design decisions.

Several limitations of the present framework should nevertheless be emphasized. First, the natural coordinates derived here make the fitted distributions approximately Gaussian, but they do not provide a mathematical guarantee of exact normality. Residual skewness can remain when a parameter has extremely weak sensitivity, when the posterior distribution is strongly truncated by parameter bounds, or when multiple local minima contribute to the fit. In such cases, the transformed analytical method should be checked against normalized Monte Carlo sampling or Bayesian posterior sampling. Second, the covariance formulation is still a local method: it assumes that the inverse problem can be adequately linearized near the optimum in the transformed space. If the sensitivity matrix is nearly singular or if the fitted parameters are so strongly coupled that the local covariance ellipsoid no longer represents the global uncertainty structure, full probabilistic sampling remains necessary. Third, the present demonstrations assume that the known input parameters can be represented by independent Gaussian uncertainties. Correlated systematic uncertainties, such as shared errors in film thickness, heat capacity, transducer properties, or laser spot size, can in principle be incorporated through a non-diagonal input covariance matrix, but this extension has not been systematically examined here. Finally, the proposed coordinates are derived from the diffusive multilayer heat-conduction equation. They are therefore not expected to apply unchanged to regimes where the governing physics differs fundamentally, such as ballistic, quasi-ballistic, hydrodynamic, or strongly non-Fourier heat transport.

Within these boundaries, the workflow is expected to apply to thermoreflectance techniques that share the same multilayer heat-diffusion kernel (Eq. (2)), including TDTR, FDTR, SDTR, SSTR, and SPS. For techniques or materials governed by fundamentally different transport equations, an analogous derivation should be performed from the corresponding governing model before applying covariance-based propagation.

When such coordinates are not obvious from the governing equations, the present framework also suggests a possible data-driven extension. The physics-derived transformations used here are conceptually related to the change-of-variables principle underlying normalizing flows [34, 35], where a non-Gaussian distribution is mapped into a simpler latent space. In future work, invertible mappings trained on Monte Carlo or Bayesian posterior samples could be used to identify approximately Gaussian coordinates for more complex inverse problems. However, such learned transformations would generally be numerical rather than closed-form, and would require dimensional constraints, cross-geometry validation, and physical interpretability checks before they could serve as robust uncertainty-propagation coordinates.

## 4. CONCLUSIONS

This work introduces a high-fidelity, computationally efficient framework for uncertainty quantification in thermoreflectance inverse problems. By pairing normalized Monte Carlo sampling with covariance-based analytical propagation in physically transformed variables, we reveal two critical insights. First, Monte Carlo analysis is not automatically reliable when the fitted parameters differ greatly in magnitude: direct fitting in the original dimensional parameter space artificially broadens the inferred distributions and overestimates uncertainty. Fitting normalized parameters relative to their optimal estimates is a practical prerequisite for trustworthy Monte Carlo estimation. Second, we show that the multilayer heat-diffusion equation dictates an intrinsic statistical geometry. The approximately Gaussian variables are the thermal resistance $1/G$ for interfaces, the geometric-mean conductivity $\sqrt{k_z k_r}$ and thermal effusivity $\sqrt{k_z C}$ for substrates, and the in-plane thermal diffusivity $k_r/C$, areal heat capacity $hC$, and cross-plane diffusion time $Ch^2/k_z$ for thin films. Once uncertainty propagation is performed in these coordinates, the assumptions underlying fast covariance-based analytical methods are substantially better satisfied, yielding results in close agreement with normalized Monte Carlo simulations across six representative multilayer systems.

The resulting framework resolves the long-standing trade-off between computational efficiency and statistical fidelity in thermoreflectance uncertainty analysis. The analytical route reduces the evaluation time from hours of repeated nonlinear optimization to a single Jacobian evaluation, while retaining the accuracy of the Monte Carlo benchmark. This efficiency gain stems from identifying the correct statistical coordinates in which the inverse problem is locally linear and approximately Gaussian. The workflow is expected to apply to any thermoreflectance modality sharing the multilayer heat-diffusion kernel of Eq. (2), and provides a template for coordinate transformations in other non-Gaussian inverse heat-conduction problems. Extensions to non-diffusive transport regimes or Bayesian frameworks with physical priors remain promising directions for future work.

## ACKNOWLEDGMENT

This work is supported by the National Natural Science Foundation of China (NSFC) through Grant No. 52376058.

Supplementary Information

# Physics-Derived Natural Coordinates for Fast Uncertainty Propagation in Multilayer Thermoreflectance

Tao Chen[1], Dihui Wang[2], Puqing Jiang[1,*]

[1]*School of Power and Energy Engineering, Huazhong University of Science and Technology, Wuhan, Hubei 430074, China*

[2]*Swanson School of Engineering, University of Pittsburgh, Pittsburgh, Pennsylvania, USA*

## S1. Numerical effect of fitting-variable normalization

Each Monte Carlo (MC) realization for the Al–GaN–AlGaN–Si case was fitted twice, once using the original dimensional parameters and once using the normalized parameters. The two fits used identical realizations of the known input parameters, identical signal perturbations, and identical physical parameter bounds. Figure S1 compares the final objective-function values obtained from the two parameterizations on a trial-by-trial basis. Normalized fitting yields a lower final objective-function value in every trial, while the corresponding root-mean-square fitting residuals remain comparable to the experimental noise level. These results confirm that the discrepancy between the two MC distributions arises from numerical convergence effects associated with poor parameter scaling.

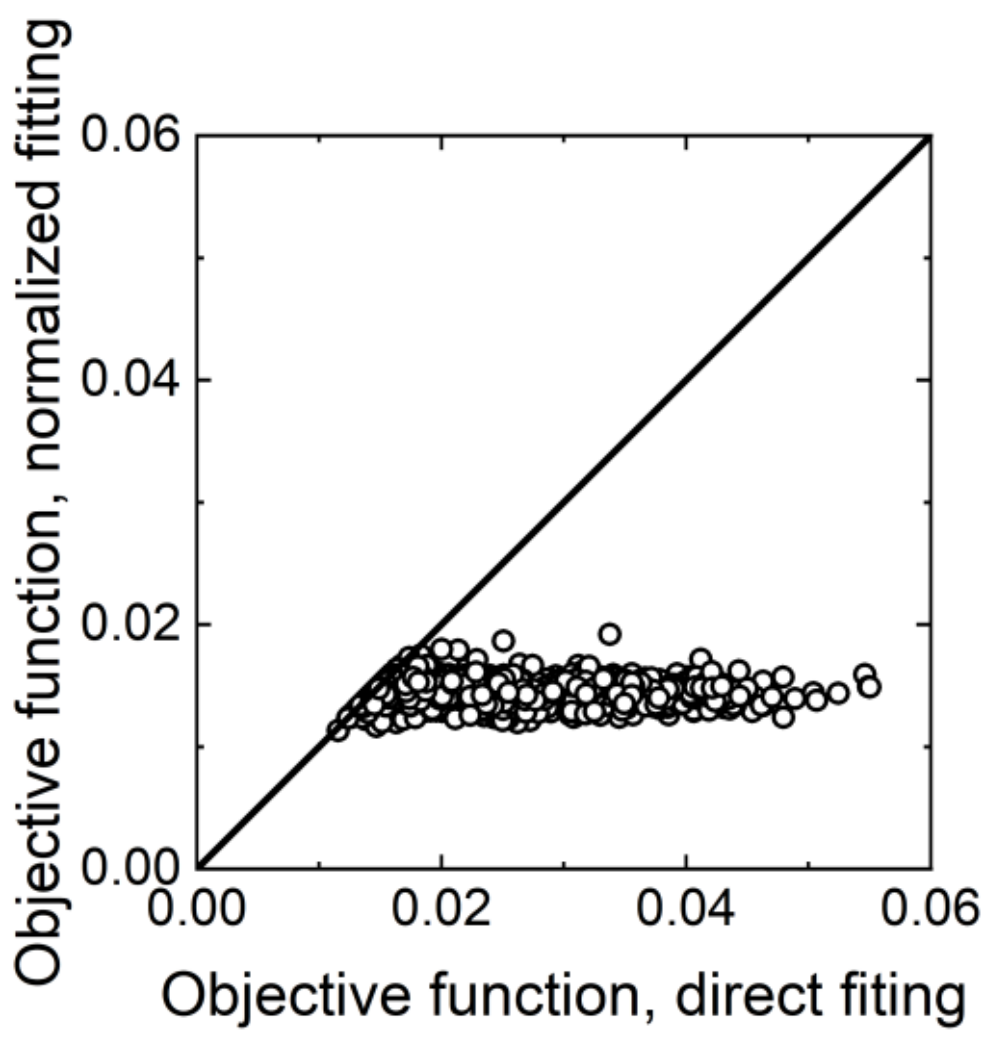

**Fig. S1.** Trial-by-trial comparison of the final objective-function values obtained from direct dimensional fitting and normalized fitting for the Al–GaN–AlGaN–Si case. Each point represents the same Monte Carlo realization fitted using the two parameterizations with identical physical bounds.

*Corresponding Author: jpq2021@hust.edu.cn (P.J.)

## S2. Joint Gaussianity assessment for the six FDTR cases

The joint distributions of the natural coordinates were evaluated using the squared Mahalanobis distances of the normalized-MC samples. Following the multivariate-normality graphical assessment based on the chi-square Q–Q plot, the empirical Mahalanobis-distance quantiles were compared with the corresponding $\chi_p^2$ quantiles, where $p$ is the dimension of the natural-coordinate vector. Figure S2 shows the Q–Q plots and the corresponding empirical and theoretical cumulative distributions for all six representative cases. Overall, the MC results follow the multivariate-Gaussian reference reasonably well, supporting the use of a multivariate Gaussian approximation for covariance-based uncertainty propagation. Some deviations are observed for the Al–GaN–AlGaN–Si case; nevertheless, the analytical 2RSD values agree with the MC benchmarks within 4.7 percentage points, indicating sufficient accuracy for the present uncertainty analysis.

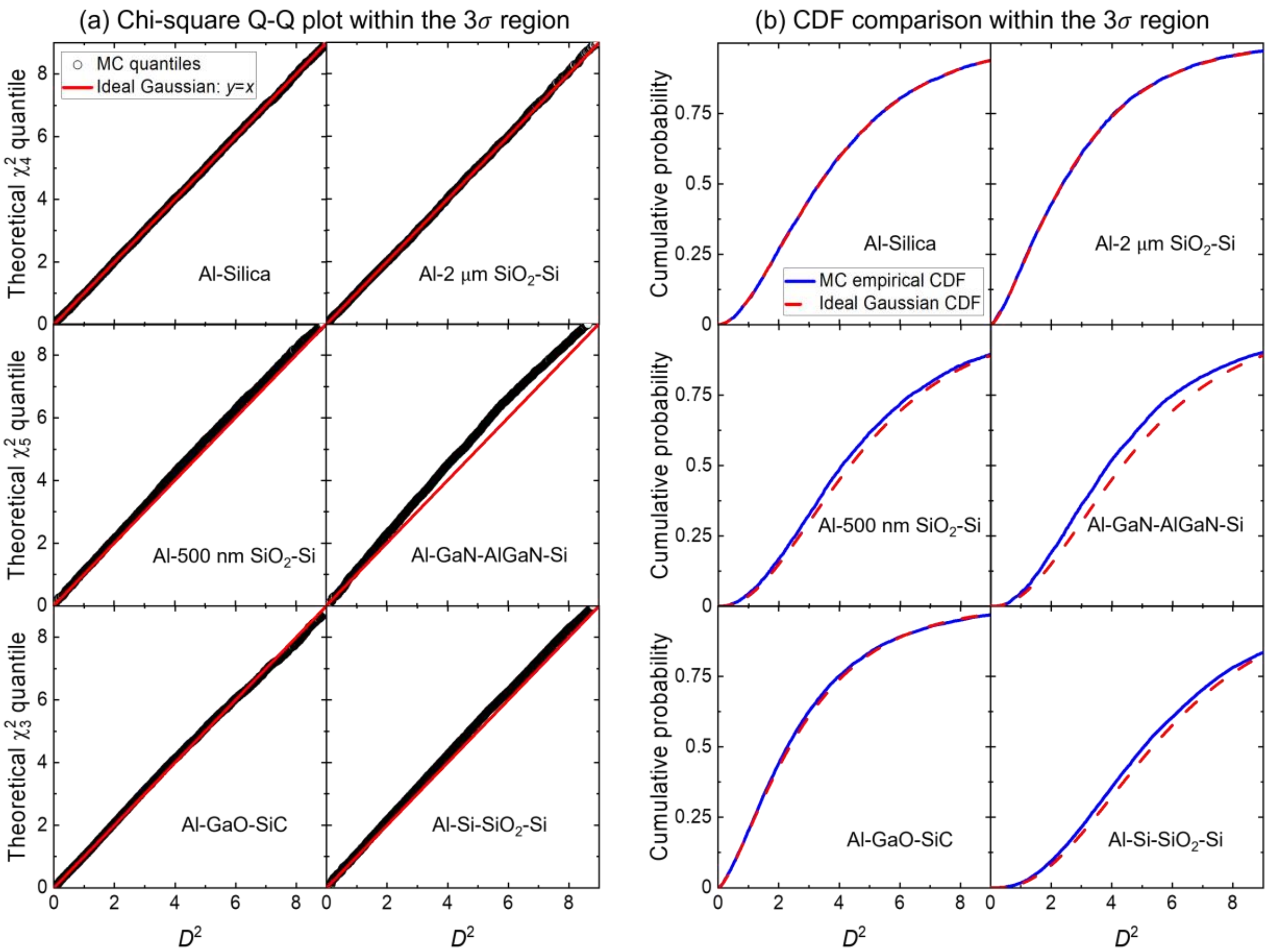

**FIG. S2.** Multivariate-Gaussian assessment of the natural-coordinate vectors obtained from normalized Monte Carlo analysis for the six representative cases. Left panels: chi-square Q–Q comparisons between the empirical squared Mahalanobis distances $D^2$ and the theoretical $\chi_p^2$ quantiles, where $p$ is the dimension of the corresponding natural-coordinate vector. Right panels: comparison of the empirical MC cumulative distributions with the ideal multivariate-Gaussian $\chi_p^2$ distributions.

**S3. Cross-modality validation of the natural coordinates**

Six additional cases were evaluated: three SPS configurations (Al/GaO/SiC, Al/2 μm $SiO_2$/Si, and Al/2 μm Si/$SiO_2$), two TDTR configurations (Al/GaO/SiC and Al/2 μm $SiO_2$/Si), and one SSTR configuration (Al/50 nm Si/$SiO_2$ measured against an Al/sapphire reference). Each case contains 1000 independently perturbed full nonlinear signal refits. For Al/GaO/SiC, the natural vector is ($1/G$, $\sqrt{k_z k_r}$, $\sqrt{k_z}$), because $C$ is fixed. For the SPS/TDTR thin-film cases, ($1/G$, $1/k_z$, $k_r$) is equivalent to interfacial resistance, cross-plane diffusion time, and in-plane diffusivity when film thickness and heat capacity are fixed.

Figures S3–S7 compare the marginal Monte Carlo distributions in the original physical coordinates with those in the corresponding natural coordinates. The same transformations identified from the multilayer heat-diffusion equation remain effective for both SPS and TDTR. The improvement is particularly evident in cases where the original $G$ or $k_z$ distributions are strongly skewed: the corresponding $1/G$ or $1/k_z$ distributions become substantially more symmetric and are well represented by Gaussian fits.

Table S1 compares the relative 2RSD uncertainties obtained from full-model Monte Carlo refitting with those predicted by local analytical covariance propagation in the natural-coordinate space. Overall, the analytical method reproduces the Monte Carlo uncertainty scale well across the SPS and TDTR configurations.

The joint distributions of the SPS and TDTR natural-coordinate vectors were further examined using squared Mahalanobis-distance diagnostics. As shown in Figs. S8 and S9, the chi-square Q–Q $R^2$values are 0.971, 0.987, and 0.995 for the three SPS cases and 0.995 and 0.994 for the two TDTR cases, respectively. The corresponding empirical radial cumulative distributions also closely follow the theoretical $\chi_p^2$references. These results indicate that the natural-coordinate covariance ellipsoid captures the dominant joint uncertainty structure across the five SPS and TDTR configurations.

For SSTR, because the method is intrinsically designed to probe the low-frequency limit, the bulk effective thermal conductivity and the thermal resistance of a thin film do not require additional validation here. We therefore focus on the in-plane thermal conductivity $k_r$ of a thin film. The target sample consists of 103-nm Al / 50-nm Si / $SiO_2$, with Al/sapphire used as the reference, and both signals are evaluated at 1 Hz. The only unknown parameter in the inverse fitting is the $k_r$ of the Si film. As shown in Fig. S10(a), although $k_r$ exhibits a relatively large uncertainty, its MC distribution remains approximately Gaussian. Because the volumetric heat capacity $C$ is fixed during the inversion, the fitted $k_r$ directly represents the natural coordinate

*Corresponding Author: jpq2021@hust.edu.cn (P.J.)

$k_r/C$. The MC analysis gives a relative 2RSD uncertainty of 55.47%, in close agreement with the analytical prediction of 54.65%. Moreover, the normal Q–Q plot yields an $R^2$ of 0.9990, further confirming the near-Gaussian behavior of this natural coordinate in SSTR.

**Table S1.** Relative 2RSD uncertainties of the fitted natural coordinates obtained from Monte Carlo refitting and local analytical covariance propagation for the SPS and TDTR cases.

| Method | Case | Natural coordinate | MC 2RSD (%) | Analytical 2RSD (%) |
|---|---|---|---|---|
| SPS | Al/GaO/SiC | $1/G_{\mathrm{GaO/SiC}}$ | 19.14 | 18.77 |
| | | $\sqrt{k_{\mathrm{z,SiC}}k_{\mathrm{r,SiC}}}$ | 8.61 | 8.39 |
| | | $\sqrt{k_{\mathrm{z,SiC}}}$ | 30.86 | 30.46 |
| | Al/2 μm $SiO_2$/Si | $1/G$ | 67.51 | 68.04 |
| | | $1/k_z$ | 7.52 | 7.68 |
| | | $k_r$ | 89.43 | 99.87 |
| | Al/2 μm Si/$SiO_2$ | $1/G$ | 32.98 | 38.63 |
| | | $1/k_z$ | 75.60 | 90.12 |
| | | $k_r$ | 6.11 | 6.53 |
| TDTR | Al/GaO/SiC | $1/G$ | 14.35 | 14.69 |
| | | $\sqrt{k_z k_r}$ | 5.26 | 5.27 |
| | | $\sqrt{k_z}$ | 13.00 | 13.14 |
| | Al/2 μm $SiO_2$/Si | $1/G$ | 60.43 | 57.58 |
| | | $1/k_z$ | 19.82 | 20.57 |
| | | $k_r$ | 66.41 | 60.42 |

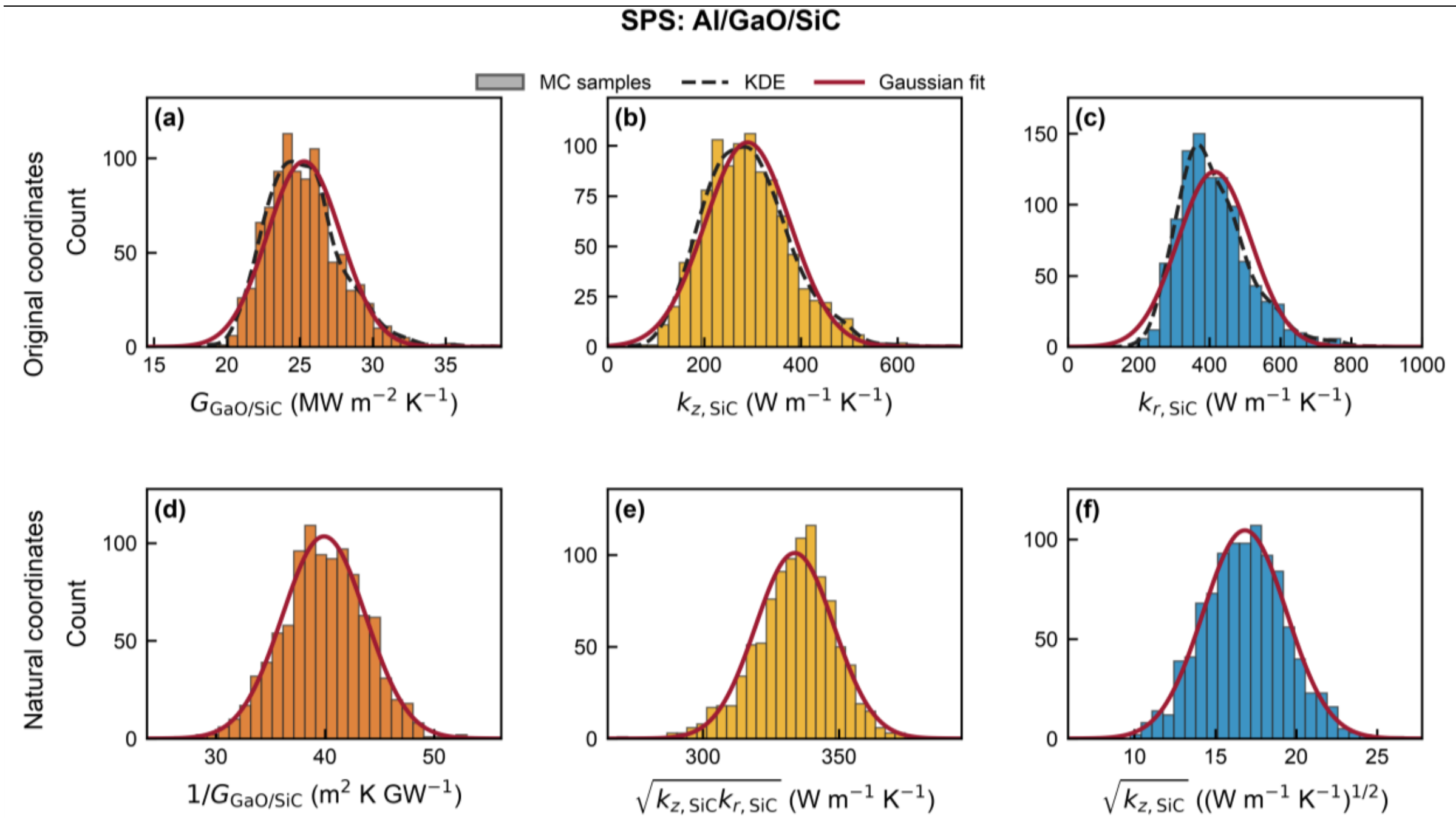

**Fig. S3.** Marginal Monte Carlo distributions for SPS Al/GaO/SiC. Top row: original coordinates ($G_{\mathrm{GaO/SiC}}$, $k_{\mathrm{z,SiC}}$, $k_{\mathrm{r,SiC}}$), with the Gaussian fit shown as a solid line and the kernel density estimate as a dashed line when distinct. Bottom row: natural coordinates ($1/G_{\mathrm{GaO/SiC}}$, $\sqrt{k_{\mathrm{z,SiC}}k_{\mathrm{r,SiC}}}$, $\sqrt{k_{\mathrm{z,SiC}}}$).

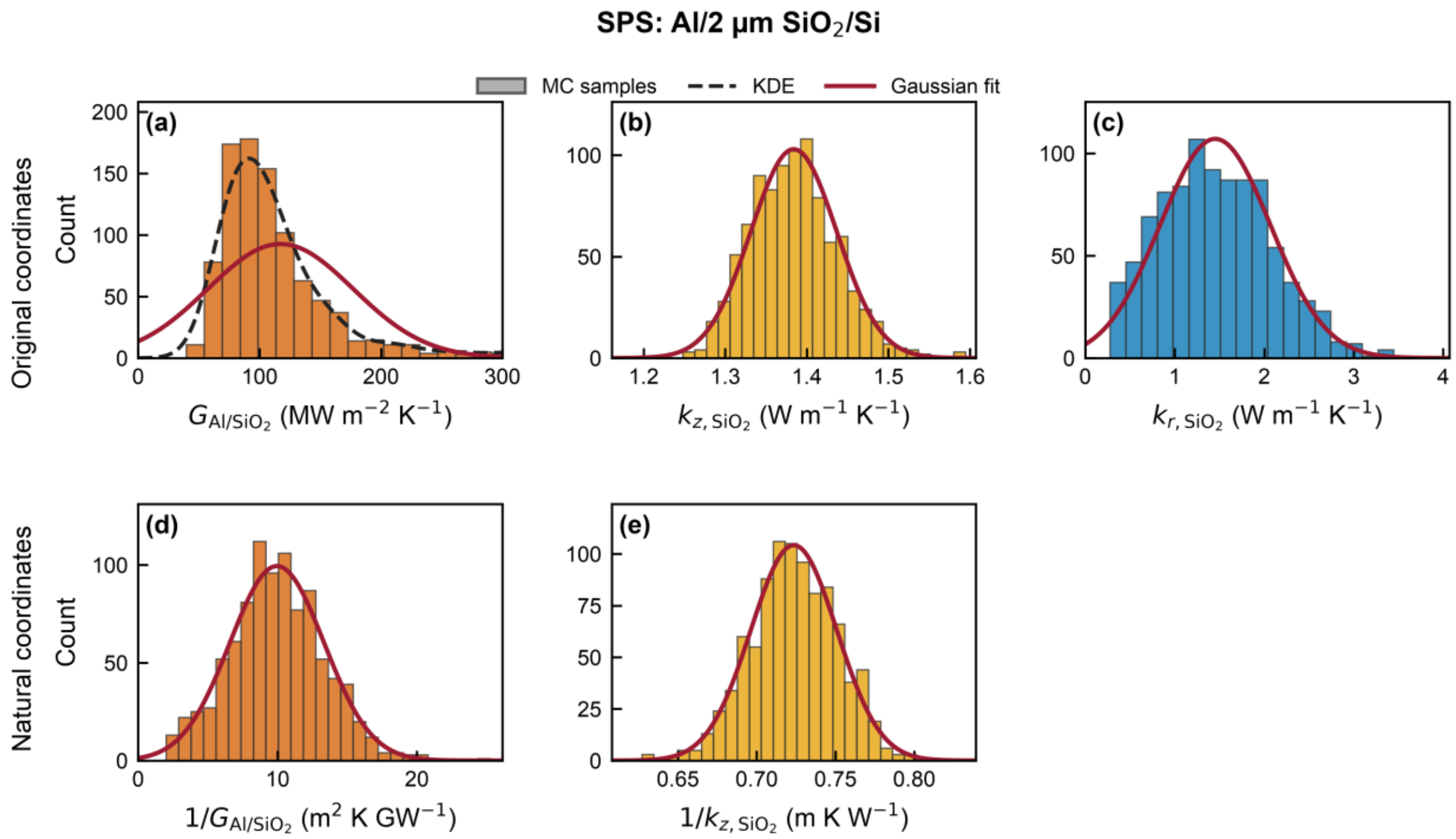

**Fig. S4.** Marginal Monte Carlo distributions for SPS Al/2 μm $SiO_2$/Si. Top row: original coordinates ($G_{Al/SiO_2}$, $k_{z,SiO_2}$, $k_{r,SiO_2}$). Bottom row: natural coordinates ($1/G_{Al/SiO_2}$, $1/k_{z,SiO_2}$, $k_{r,SiO_2}$). The unchanged $k_r$ coordinate is not repeated in the lower row.

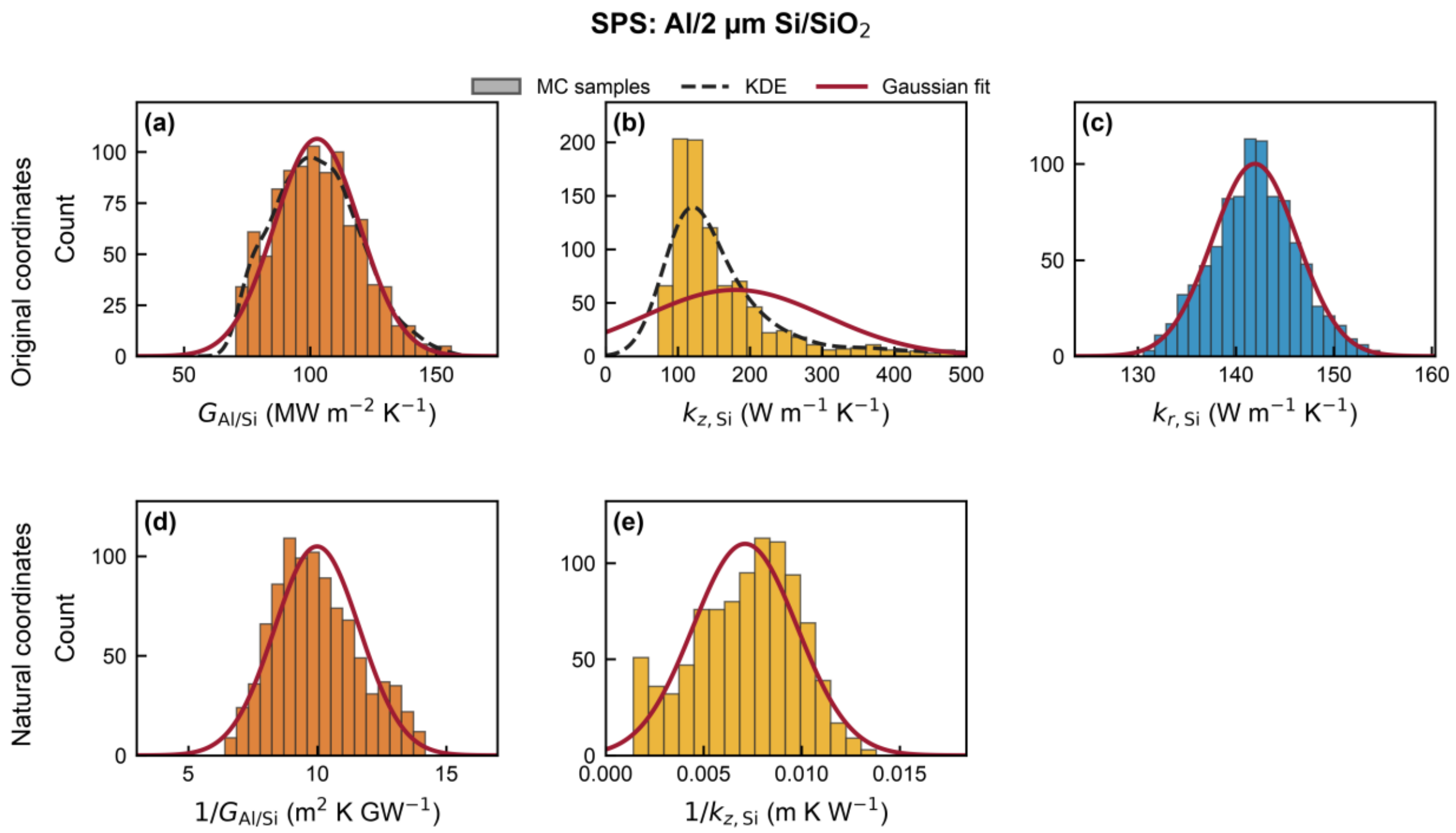

**Fig. S5.** Marginal Monte Carlo distributions for SPS Al/2 μm Si/$SiO_2$. Top row: original coordinates ($G_{Al/Si}$, $k_{z,Si}$, $k_{r,Si}$). Bottom row: natural coordinates ($1/G_{Al/Si}$, $1/k_{z,Si}$, $k_{r,Si}$). The unchanged $k_r$ coordinate is not repeated in the lower row.

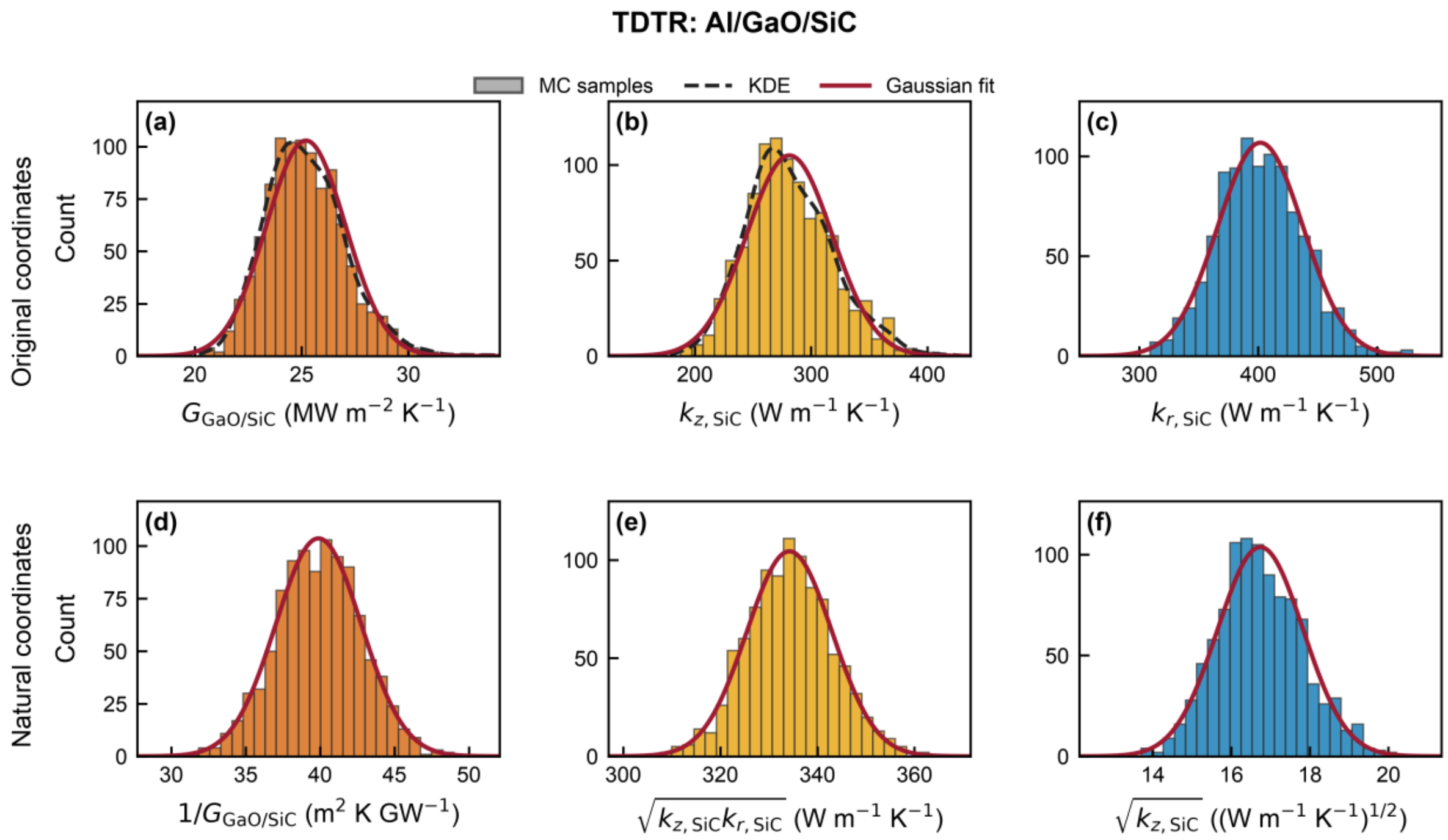

**Fig. S6.** Marginal Monte Carlo distributions for TDTR Al/GaO/SiC. Top row: original coordinates ($G_{\mathrm{GaO/SiC}}$, $k_{\mathrm{z,SiC}}$, $k_{\mathrm{r,SiC}}$). Bottom row: natural coordinates ($1/G_{\mathrm{GaO/SiC}}$, $\sqrt{k_{\mathrm{z,SiC}}k_{\mathrm{r,SiC}}}$, $\sqrt{k_{\mathrm{z,SiC}}}$).

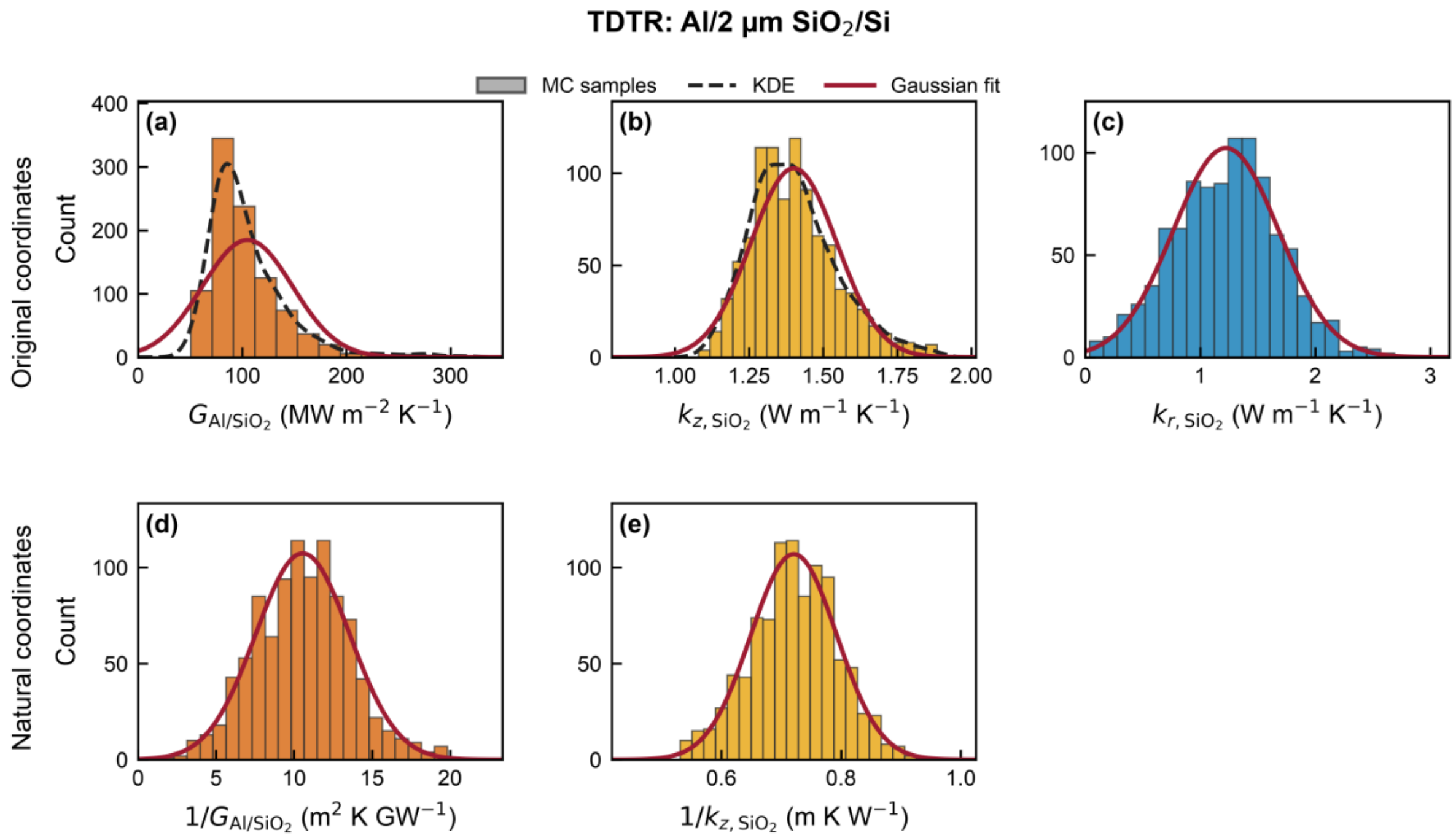

**Fig. S7.** Marginal Monte Carlo distributions for TDTR Al/2 μm $SiO_2$/Si. Top row: physical coordinates ($G_{\mathrm{Al/SiO_2}}$, $k_{\mathrm{z,SiO_2}}$, $k_{\mathrm{r,SiO_2}}$). Bottom row: natural coordinates ($1/G_{\mathrm{Al/SiO_2}}$, $1/k_{\mathrm{z,SiO_2}}$, $k_{\mathrm{r,SiO_2}}$). The unchanged $k_r$ coordinate is not repeated in the lower row.

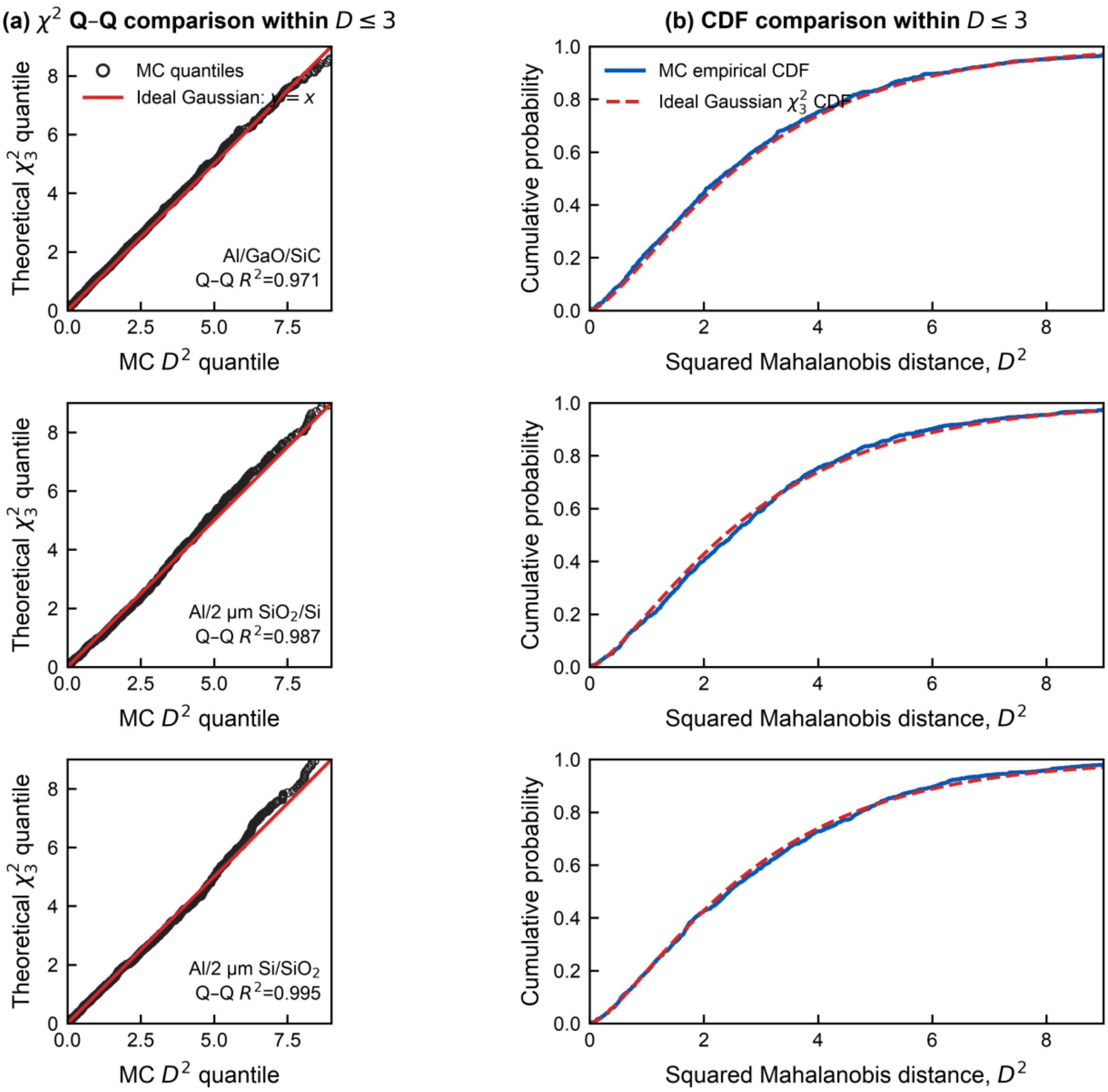

**Fig. S8.** Full-dimensional natural-coordinate diagnostics for the three SPS cases. Left: theoretical $\chi_3^2$ quantiles versus empirical squared-Mahalanobis-distance quantiles from the Monte Carlo results. Right: empirical and ideal radial cumulative distributions. Q–Q $R^2$ values are reported within the left panels.

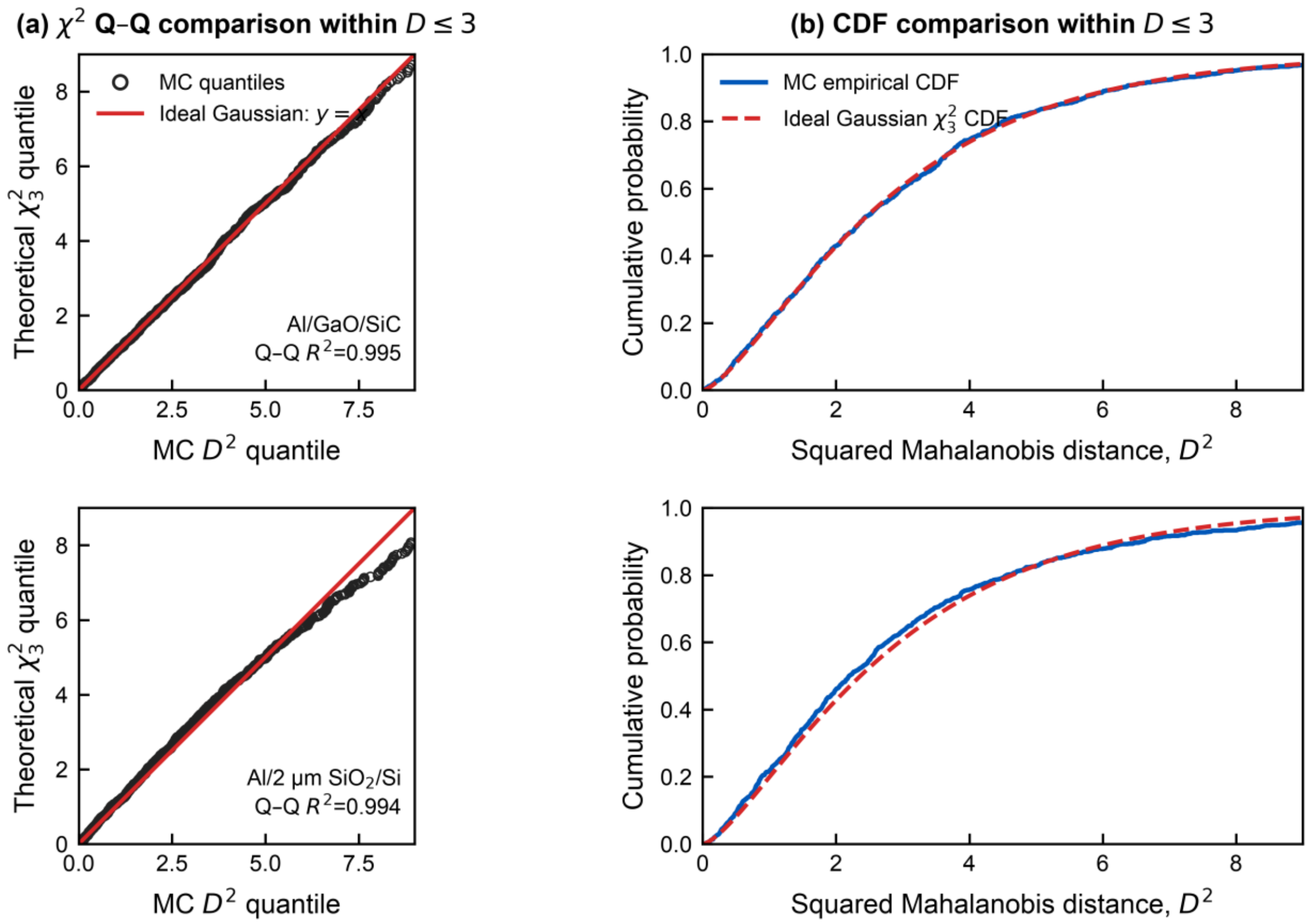

**Fig. S9.** Full-dimensional natural-coordinate diagnostics for the two TDTR cases. Left: theoretical $\chi^2_3$ quantiles versus empirical squared-Mahalanobis-distance quantiles from the Monte Carlo results. Right: empirical and ideal radial cumulative distributions. Q–Q $R^2$ values are reported within the left panels.

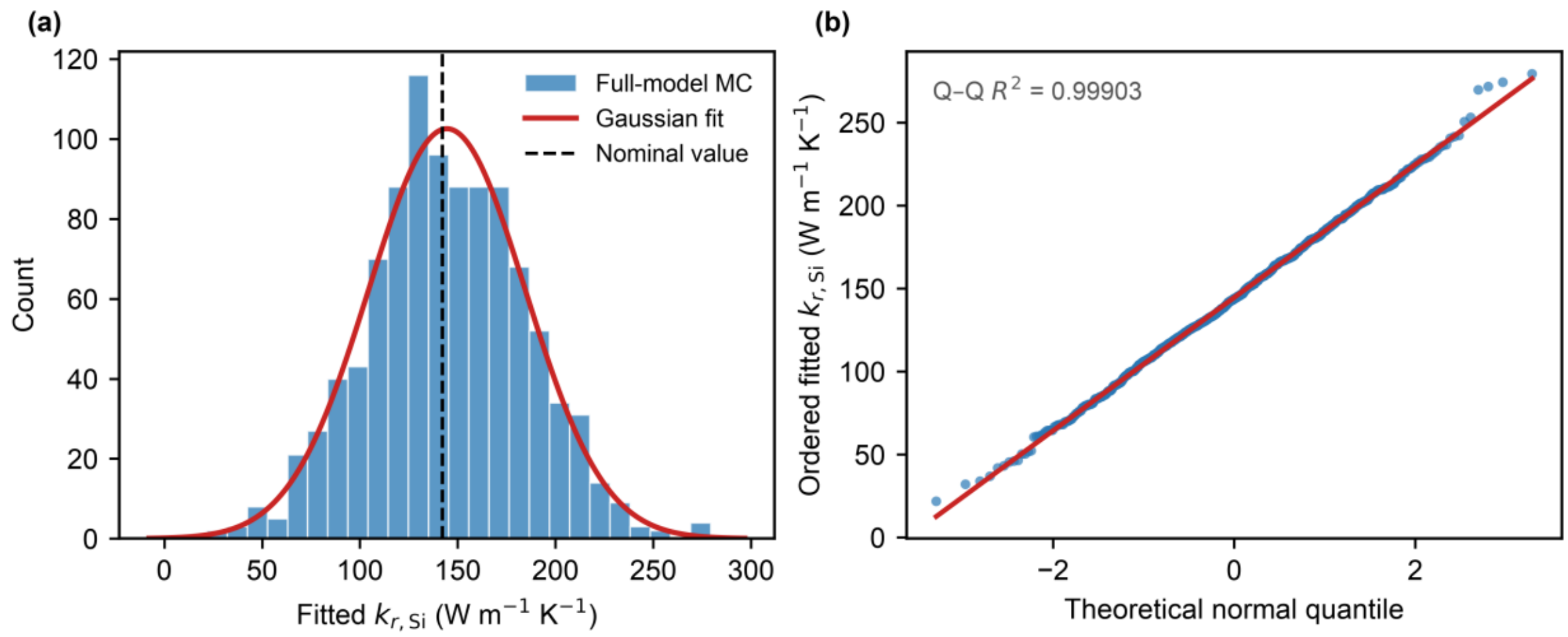

**Fig. S10.** Monte Carlo validation for SSTR Al/50 nm Si/$SiO_2$ measured against an Al/sapphire reference. (a) Distribution of the fitted Si-film $k_r$ and its Gaussian fit. (b) Normal Q–Q comparison.